\documentclass[useAMS,usenatbib]{mn2e}
\usepackage{graphicx}
\usepackage{amsmath}
\usepackage{subfigure}
\usepackage{amssymb}
\usepackage{tabularx}
\usepackage{natbib}
\usepackage{tensor}
\usepackage[flushleft]{threeparttable}
\usepackage{booktabs,caption,fixltx2e}

\def\bge{\begin{equation}}
\def\ede{\end{equation}}

\newcommand{\erf}{\mathrm{erf}\,}

\title[The migratory origin of NSCs]
  {The Globular Cluster Migratory Origin of Nuclear Star Clusters}
\author[M. Arca-Sedda and R. Capuzzo-Dolcetta]
  {M.~Arca-Sedda$^1$$^,$$^2$,
   R.~Capuzzo-Dolcetta$^1$ \\
  \\
  $^1$Sapienza-Universit\`{a} di Roma, P.le A. Moro 5, I-00165 Rome, Italy\\
$^2$University of Tor Vergata, Via O. Raimondo 18, I-00173 Rome, Italy}
\date{Released 2014 Xxxxx XX}

\pagerange{\pageref{firstpage}--\pageref{lastpage}} \pubyear{2002}

\def\LaTeX{L\kern-.36em\raise.3ex\hbox{a}\kern-.15em
    T\kern-.1667em\lower.7ex\hbox{E}\kern-.125emX}

\begin{document}

\label{firstpage}

\maketitle

\begin{abstract}
Nuclear Star Clusters (NSCs) are often present in spiral galaxies as well as resolved Stellar Nuclei (SNi) in elliptical galaxies centres. Ever growing observational data indicate the existence of correlations between the properties of these very dense central star aggregates  and those of host galaxies, which constitute a significant constraint for the validity of theoretical models of their origin and formation. 
In the framework of the well known 'migratory and merger' model for NSC and SN formation, in this paper we obtain, first, by a simple argument the expected scaling of the NSC/SN mass with both time and parent galaxy velocity dispersion in the case of dynamical friction as dominant effect on the globular cluster system evolution. This generalizes previous results by \cite{TrOsSp} and is in good agreement with available observational data showing a shallow correlation between NSC/SN mass and galactic bulge velocity dispersion.
Moreover, we give statistical relevance to predictions of this formation model, obtaining a set of parameters to correlate with the galactic host parameters. We find that the correlations between the masses of NSCs in the migratory model and the global properties of the hosts reproduce quite well the observed correlations, supporting the validity of the migratory-merger model. In particular, one important result is the flattening or even decrease of the value of the NSC/SN mass obtained by the merger model as function of the galaxy mass for high values of the galactic mass, i.e. $\gtrsim 3\times 10^{11}$M$_\odot$, in agreement with some growing observational evidence.
\end{abstract}


\section{Introduction}
Due to the ever growing quantity of high resolution data, in the last few years great interest has been focused on the central region of galaxies where various phenomena co-exist.

Thanks to the high resolution images provided by the Hubble Space Telescope, it is clear, nowadays, that the nuclei of the majority of both elliptical and early type spiral galaxies  ($M>10^{10}$M$_\odot$) harbour massive or supermassive black holes (SMBHs), whose masses, $M_{BH}$, range in the $10^6-10^9$M$_\odot$ interval and may be up to $\sim 10^{10}$ M$_\odot$ as in the case of the SMBH in NGC1277  \citep{VdB}.
In some cases, the central SMBH is surrounded by a massive, very compact, star cluster commonly referred as Nuclear Star Cluster (NSC).

NSCs are observed in galaxies of every type of the Hubble sequence \citep{bkr,cote06} and their modes of formation and evolution are still under debate. In the case of elliptical galaxy hosts, the nuclear clusters are also referred to as 'resolved stellar nuclei'. For the sake of this paper we will refer to NSCs or resolved stellar nuclei indifferently.
NSCs are sited at the photometric and kinematic centre of the host galaxy, i.e. at the bottom of the potential well
\citep{BKR02,Neum}. This is likely connected to a peculiar formation history. As a matter of fact, all NSCs contain an old stellar population ($age>1$ Gyr) and most of them show, also, the presence of a young population, with ages below $100$ Myr \citep{rossa,Seth,Neum}. 

NSCs are bright (about 4 mag brighter than ordinary globular clusters), massive objects ($10^6-10^7$ M$_\odot$), very dense and with a half-light radius of $2-5$ pc. Their small sizes and large masses make them the densest stellar systems in the Universe  \citep{Neum12}.

The relation between NSCs and SMBHs is poorly known; they seem to be two 'faces of the same coin', constituting central massive objects (CMOs) whose actual presence depends on the host mass: galaxies with mass above  $10^{10}$ M$_\odot$ usually host an SMBH while lighter galaxies have, instead, a well resolved central star cluster (an NSC). Moreover, a transition region exists for galaxies with mass between $10^8$ and $10^{10}$ M$_\odot$ in which both the objects co-exist \citep{BKR02,Bkr09,Graham}.

With regard to the lack of evidence of NSCs in high mass ($> 10^{11}$ M$_\odot$) galaxies, one possible explanation is the formation of giant ellipticals through merging of smaller galaxies \citep{Merri}. 
Quantitatively speaking, \cite{bekkiGr} simulations showed that if the two colliding galaxies host MBHs, a black hole binary (BHB) could form which heats up the resulting stellar nucleus causing its progressive evaporation. This process can destroy the 
super cluster, shaping significantly the density profile of the merger product and leaving behind a BHB that shrinks due to gravitational wave emission leading eventually to a SMBH.
Another possibility is that in the early phase of the galaxy life, an initial NSC could be the seed for the BH birth as suggested first by \cite{Dolc93} and later by \cite{Neum} and \cite*{Gne14}.

Recently, a number of researchers studied the existence of  scaling relations between NSCs and their galactic hosts; similar studies have already been done, seeking for scaling relations between SMBHs and their hosts. However, it is still unclear the robustness of the NSC-galaxy relations and whether these relations are linked to those between SMBHs and the hosts. 

For instance, \cite{frrs} showed that the NSC mass ($M_{\rm NSC}$) vs galaxy velocity dispersion ($\sigma$) relation is roughly the same of that observed for SMBHs. On the other hand, more recent studies \citep*{LGH,Graham} claim that the $M_{\rm NSC}-\sigma$ relation is shallower than for SMBHs, $M_{\rm NSC}\propto \sigma^{1.5}$. Moreover, it has been shown that while SMBH masses correlate with the galaxy mass, the NSC masses correlate better with the bulge mass \citep{ERWGD}.

At present, two are the most credited frameworks for the NSC formation.

One scenario refers to the so called (dissipational) 'in-situ model' \citep*{King03,King05,Mil04,Beketal06}. 

According to this model, an injection of gas in the central region of a galaxy hosting a 'seed' black hole could lead to the formation of a NSC if the typical crossing time of the parental galaxy is shorter than the so-called 'Salpeter time', which is the time-scale over which the central BH can grow by accretion \citep*{nayakshin}. 

Another (dissipationless) scenario invokes the action of the dynamical friction process which makes massive globular clusters (GCs) sink toward the centre of the host galaxy \citep*{TrOsSp,Pes92,Dolc93}. Their subsequent merging leads to a super star cluster with characteristics indistinguishable from those of an NSC \citep{DoMioA,DoMioB}. This scenario is often referred to as infall-merger scenario or migratory-merger model.

Both the theories above encounter some troubles in explaining completely the NSC formation. According to some qualitative considerations, the in situ model would predict too  massive NSCs, while a possible problem for the GCs infall model is that it would give lighter NSCs than observed \citep{LGH}. \cite{Hart11} say that mergers of star clusters are able to produce a wide variety of observed properties, including densities, structural scaling relations, shapes (including the presence of young discs) and even rapid rotation, nonetheless claim that some kinematical properties of observed NSCs are hardly compatible with merger models. They suggest the need of a 50$\%$  of gas in the overall scheme.

\cite{Turetal12} referring to their Fornax ACS survey and to some speculative considerations conclude that, for galaxies and nuclei in their sample, the infall formation mechanism is the more likely for low to intermediate mass galaxies while for more massive ones 
accretion triggered by mergers, accretions, and tidal torques is likely to dominate. The two mechanisms smoothly vanish their efficiency on the intermediate mass galactic range, and they indeed provide some evidence of "hybrid nuclei" which could be the result of parallelly acting formation mechanisms.

While the in-situ model has remained, so far, almost speculative and difficult to constrain to available observations, several authors provided detailed numerical tests for the GC merger scenario starting from the original idea in \cite{Dolc93}. The first simulations were done by \citep{DoMioA} and \citep{DoMioB} in galaxy models without massive black holes and stellar discs; \citep{bekki10} studied the role of stellar discs and \cite{AMB} the role of a central galactic MBH.

In particular, \cite{AMB} made a full $N$-body simulation of the decay and merging of 12 GCs in a Milky Way model accounting for the presence of the Sgr A* $4\times 10^6$ M$_\odot$ central black hole, obtaining an NSC that has global properties fully consistent with those observed in the nucleus of our galaxy.
One recent work by \cite{PM14} presents the same merger simulations performed in \cite{AMB} with the inclusion of different stellar populations in the various infalling globular clusters, and shows that infalling clusters can produce thick flattened structures with varied orientations, possibly related to 'disky' structures that are observed in galactic nuclei and clusters (see \cite{MBP13} for a discussion of the evolution of such discs).

On another side, \cite{Ant13}, by mean of a semyanalytical model, made some comparisons among the expexcted result of a merger scenario for the NSC formation and some scaling laws.

The aim of this paper is to check in a more complete and extensive way the reliability of the infall-merger model for the NSC formation. To reach this aim we build  'theoretical' scaling laws connecting NSC properties with those of the galactic hosts in a synthetic modelization of the global evolution of a Globular Cluster System (GCS) in a galaxy, considering the dynamical friction and tidal disruption as evolutionary engines. These scaling laws are to be compared with those observationally obtained.

The paper is organized as follows: in Section \ref{dynfri} the role of dynamical friction is discussed as well as the way we modeled galaxies and their GCS; in the same section the sample of data used for the comparison with observation is presented;
Section \ref{appro} presents two different theoretical modelizations of the NSC growth in galaxies;
in Section \ref{laws} we provide a set of 'theoretical' scaling laws which connect NSCs with their hosts and compare them with the observed laws; in Section \ref{tid}, instead, we take into account the effect of tidal disruption of GCs on the resulting NSC mass. Finally, Section \ref{end} is devoted to a summary of the main results, providing some general remarks such to draw conclusions.

\section{The Globular Cluster infall scenario}
\label{dynfri}

The formation of a compact nucleus in the centre of a galaxy through the orbital decay of globular clusters has been discussed, first, by \cite{TrOsSp}. Working on a model of the M31 galaxy, they demonstrated that the efficiency of the dynamical friction mechanism could provide an amount of matter sufficient to form a compact nucleus of $10^7-10^8 $M$_\odot$ in the centre of this galaxy. \cite{Dolc93} turned out the importance of considering the tidal disruption of the clusters as a competitive process that tunes the effect of dynamical friction.
Before approaching in a deeper way the theme, we now give a relevant analytical support to the idea that a NSC can grow in the centre of a galaxy by mean of GC decay in the innermost region via dynamical friction braking.

\subsection{A preliminary, relevant scaling result}
\label{teo}

A direct and easy way to obtain a scaling between the mass accumulated to the galactic centre and the background velocity dispersion is based on the assumption that the galaxy has the mass density of a singular isothermal sphere 

\bge
\rho(r)=\frac{v_{\rm c}^2}{4\pi G r^2},
\ede

where $G$ is the gravitational constant, $r$ is the galactocentric distance and $v_{\rm c}$ is the circular velocity, constant with radius and related to the velocity dispersion, $\sigma$, by $v_{\rm c}=\sqrt{2}\sigma$.

Approximating the motion of the test mass, $M$, as a decreasing energy sequence of circular motions, the evolution equation of the modulus of the orbital angular momentum per unit mass, $\textbf{L}=\textbf{r}\wedge \textbf{v}$, is

\begin{equation} 
\dot{L} = \dot{r}v_{\rm c}=-r \frac{F_{\rm df}}{M},
\end{equation}

where $F_{\rm df}$ is the absolute value of the dynamical friction force exerted by the galaxy on the test object, which, using the Chandrasekhar's formula in its local approximation, is given by

\begin{equation} 
F_{\rm df}(r) = \frac{4\pi  G^2\ln \Lambda\rho M^2}{{v_{\rm c}}^2} f(X),
\end{equation}

with

\begin{equation} 
f(X)=\erf(X)-\frac{2X}{\sqrt{\pi}},
\end{equation}

where $X\equiv v_{\rm c}/(\sqrt{2}\sigma)$ ($X=1$ for a singular isothermal sphere), $\erf(X)$ is the usual error function and $\Lambda$ is the Coulomb logarithm. 
The time evolution of the radius $r(t)$ of the nearly circular orbit of the test mass under the previous assumptions is thus governed by the differential equation 

\begin{equation}
\dot{r} =-\frac{G\ln \Lambda  Mf(1)}{\sqrt{2}\sigma}\frac{1}{r},
\end{equation}

where $f(1) \simeq 0.4276$, which, with the initial condition $r(0)=r_0$, is easily integrated to give

\begin{equation}
r(t)^2=r^2_0 -\frac{0.6047G\ln \Lambda M}{\sigma}t
\end{equation}

and leads to  $T= \sigma r^2_0/(0.6047G\ln \Lambda M)$ as fully decay time ($r(T)=0$) of the object of mass $M$ initially moving on the circular orbit of radius $r_0$..

Given, for the GC population, a density distribution in the form of a power-law 

\begin{equation}
\rho_{\rm GCS}(r)=Ar^\alpha
\end{equation}

where $\alpha >-3$ is an a priori free parameter and $A$ is a normalization constant constrained to give the total mass of the GCS, $M_{\rm GCS}$:

\begin{equation}
A = \frac{\alpha+3}{4\pi}\frac{M_{\rm GCS}}{R^{\alpha+3}}.
\end{equation}

Assuming the total GCS mass to be a fraction $f<1$ of the galactic mass $M_{\rm g}$, 

\begin{equation}
M_{\rm GCS} = fM_{\rm g} = f \frac{2\sigma^2}{G}R,
\end{equation}

$A$ turns out to be a function of the galactic velocity dispersion and radius

\begin{equation}
A= f\frac{\alpha+3}{2\pi G}\frac{\sigma^2}{R^{\alpha+2}}.
\end{equation}

If the GCs dynamically decayed to the galactic centre go to grow a nucleus therein, the value of the nucleus mass at any age, $t$, of the galaxy can be obtained by mean of 
$\overline r \equiv \overline r (t) = \sqrt{ 0.6047G\ln \Lambda Mt/\sigma}$, which is the maximum radius of the GC circular orbit decayed to the centre within time $t$.
Consequently, $M_{\rm n}(t)$, is simply 

\begin{align}
M_{\rm n}(t) &= \frac{4\pi A}{\alpha+3}\overline r^{\alpha+3}= \nonumber \\  
       &= f\frac{2}{G}\frac{\sigma^2}{R^{\alpha+2}}\overline r^{\alpha+3} =  \nonumber \\ 
       &=  f\frac{2}{G}{\left(0.6047G\ln \Lambda M\right)} ^{\alpha+3}t^{\frac{\alpha +3}{2}}\frac{\sigma^{\frac{1-\alpha}{2}}}{R^{\alpha+2}},
\label{mnt}
\end{align}

for $t\leq \sigma R^2/(0.6047G\ln \Lambda M)$, while  $M_{\rm n}(t)$ saturates to $M_{\rm GCS}$ at $t=\sigma R^2/(0.6047G\ln \Lambda M)$. 

Note that Equation \ref{mnt} is independent of the galactic radius $R$ and reduces to the $M_{\rm n} \propto \sigma^{3/2}$ relation obtained by \cite{TrOsSp} in the case of $\alpha=-2$, i.e. for GC distributed the same way as the galactic isothermal background. This is the only case where the dependence on $R$ cancels out.
If, instead, $\alpha \neq -2$ the dependence of $M_{\rm n}(t)$ on $\sigma$ becomes, in the assumption of a virial relation between galactic $R$ and $M_{\rm g}$ ($R \propto M_{\rm g}/\sigma^2$)

\begin{equation}
M_{\rm n}(t) \propto \frac{\sigma^\frac{9+3\alpha}{2}}{M_{\rm g}^{\alpha+2}},
\end{equation}

which corresponds, assuming a constant $M_g$, to a slope in the range from $0$ of the steeper ($\alpha = -3$)  GCS radial distribution to $9/2$ of the flat ($\alpha=0$) distribution.

The relevant result here is that the slope of the $M_{\rm n}-\sigma$ relation in the regime of dynamical friction dominated infall process is expected to have an upper bound in any case smaller than that of the $M_{\rm BH} - \sigma$ relation.

\subsection{The data sample}
\label{model}

The aim of this work is to show that the dry merger scenario can reproduce the correlations of observed NSCs with their hosts in a wide range of galaxy masses. 
	To reach this aim, we need three important ingredients: i) a robust data base to compare our results with real observations of NSCs and their hosts, ii) a reliable treatment of the dynamical friction and tidal disruption processes and iii) a detailed model for the host galaxies to reproduce the environment in which GCs evolve.

The data base for the purposes of this work has been extracted from three different papers. The first (\cite{ERWGD}, hereafter EG12), combines data coming from different works covering galaxies of the Hubble types S0-Sm; on another side, \cite{LGH} (hereafter LKB12) provide data for $51$ early type galaxies in the Advanced Camera Virgo Cluster Survey \citep{cote04}; finally, we considered data given in \cite{scot} (hereafter SG13) which is a collection of data from earlier works. 

At the end, we gathered a total sample of $112$ galaxies covering a wide range of Hubble types
which contains several structural parameters of each galaxy such as mass, effective radius, velocity dispersion, and of the NSC masses.

To evaluate reliably the dynamical friction braking of GCs in their host galaxies, we have to assume galactic density profiles.
As first approximation, we modeled galaxies as spherically symmetric distributions in the form of Dehnen's spheres whose density is:

\begin{equation}
\rho_{\gamma}(r)=\frac{\rho_{\gamma 0}}{\left(r/R_{\rm g}\right)^{\gamma}\left(r/R_{\rm g}+1\right)^{4-\gamma}},
\label{Dhn}
\end{equation}

where $\gamma \geq 0$ and $\rho_{\gamma_0}$ is linked by

\begin{equation} \nonumber
\rho_{\gamma 0}=\frac{(3-\gamma)M_{\rm g}}{4\pi {R_{\rm g}}^3},
\end{equation}

to the total mass of the galaxy, $M_{\rm g}$ and to  its length scale, $R_{\rm g}$. 

Generally, galaxies fainter than $M_V\sim -20.5$ show steep surface luminosity profiles $I(R)\propto R^{-\Gamma}$ with slope $\Gamma>0.5$ ('power-law' galaxies), while brighter galaxies show less pronounced cusps ('core' galaxies with $\Gamma<0.3$) \citep{Lauer,Merri}.  
The two slopes, $\Gamma$ and $\gamma$, are linked by the relation $\gamma \sim 1-\Gamma$ in the case $\gamma<1$ \citep{Deh93}. 
Consequently, for each galaxy mass $M_{\rm g}$, the $\gamma$ exponent is randomly chosen in the range $0 - 0.5$ for $M_{\rm g}<10^{10}$ M$_\odot$, and in the range $0.5 - 1$ for $M_{\rm g}>10^{10}$ M$_\odot$. On the other hand, since the slope of the $\rho(r)$ depends critically on the surface brightness profile and may vary from different kind of galaxies, we allowed also the $\gamma$ parameter to vary in a more general way, i.e. extracting it randomly between $0-2$ for each galaxy model, finding not significant changes in our results. For this reason, we decide to choose $\gamma$ in the two ranges explained above, in agreement with the fact that brighter galaxies seems to have flatter surface brightness profiles.

It is relevant noting that the validity of 'true' cuspidal density profile models to describe the matter distribution of galaxies has been questioned by 
many authors that claim that the core-S{\'e}rsic profiles are better suited to describe the innermost ($3$ to $10$ arcsecs) regions of early type galaxies \citep{Graham04,Dullo12}.  

However, \cite{Lauer} showed that power-law and core galaxies show more or less the same steepness of the density profiles in their outer regions, with small changes of the slope of the surface brightness profiles. Since we are interested in the study of the dynamics of stellar clusters on a relatively large scale of the galaxy and not in its innermost region, where the investigation required a specific, more accurate modelling which is out of our scopes, we consider the simple $\gamma$ density profiles as appropriate for our purposes.
This choice allows us to use the results on dynamical friction recently obtained for cuspy Dehnen's profile \citep{ACD14}. 

Actually, Dehnen's profiles have a central cusp, and it has been demonstrated that the in density cusps the classical Chandrasekhar dynamical friction formula fails (see for instance \cite{RCD05,Spurz,AntMer12}). In particular, both \cite{AntMer12} and \cite{ACD14} found that dynamical friction braking is reduced in the vicinity of a MBH.
To overcome this problem, \cite{ACD14} provided a formulation for the dynamical friction process which is valid in cuspy galaxies, giving a useful fitting expression for the dynamical friction timescale (\cite{ACD14}, Equation 21):

\begin{equation}
\tau_{df}=\tau_0(2-\gamma)(4.93-3.93e)\left(\frac{M}{M_{\rm g}}\right)^{-0.67}\left(\frac{r}{R_{\rm g}}\right)^{1.76},
\label{tdf}
\end{equation}

where $e$ is the orbital eccentricity of the text object of mass $M$ and $\tau_0$ is a normalization factor whose value is given by:

\begin{equation} \nonumber
\tau_0 (\mathrm{Myr}) = 0.3 \sqrt{\frac{R_{\rm g}^3}{M_{{\rm g},11}}},
\end{equation}

where $R_{\rm g}$ (in $\mathrm{kpc}$) is the scale radius and $M_{{\rm g},11}$  the total mass of the $\gamma$ model galaxy in unit of $10^{11}\rm M_\odot$. 
Equation \ref{tdf} gives the dynamical friction decay time for a GC of mass $M$ initially moving on an orbit of eccentricity $0\leq e \leq 1$ in a spherical galaxy of mass $M_{\rm g}$ and length scale $R_{\rm g}$.

Since $\tau_{df}$ depends, other than on $e$,  on $M_{\rm g}$, $R_{\rm g}$ and $\gamma$, to a good estimate of the dynamical friction time reliable values of these three parameters are needed.
In other words, it is important providing reliable models of the parent galaxy to ensure that the environment where dynamical friction acts is well reproduced.
The simplest way to produce such  model environments is via linking the parameters needed to establish the theoretical model to the observable quantities. 
As example, the LKB12 data sample contains the mass, $M_{\rm g}$, and effective radius, $R_{\rm e}$, which is the radius containing half of the total light. This parameter is important because it can be connected with the scale radius of the theoretical $\gamma$ model, $R_{\rm g}$.  However, the data sample contains a limited number of galaxies in the range $10^8-10^{10}$M$_\odot$ and does not provide data for heavier galaxies, hence only a small fraction of the total range of galaxy masses could be investigated and, moreover, not always both the $M_g$ and $R_e$ values are available. 

To overcome these limitations, we need to extend with some proper extrapolations to a range of galaxies covering a wider, $10^8-10^{12}$M$_\odot$, mass interval. 
To this aim, we used data in LKB12 to correlate the galaxy mass with the effective radius and found a good fitting formula linking these two quantities as:

\begin{equation}
R_{\rm e} ({\rm kpc}) = 1.78  {M_{{\rm g},11}}^{0.14},
\label{ReM}
\end{equation}

where $M_{\rm g}$ is in units of $10^{11}$M$_\odot$. Then, given the galaxy mass we can evaluate its own effective radius, which is an astrophysical observable, and, finally, we get the scale length $R_g$ by mean of these two relations  \citep{Deh93}:

\begin{equation}
	R_{\rm e}=3R_{\rm h}/4,
	\label{effec}
\end{equation}

\begin{equation}
	R_{\rm h}=\frac{R_{\rm g}}{\left[2^{1/(3-\gamma)}-1\right]}.
	\label{RH}
\end{equation}

the first one being valid in the range of $\gamma$ considered in this paper.

Finally, the scale length $R_g$ is related to $\gamma$ and $M_g$ by this relation

\begin{equation}
R_{\rm g}(\mathrm{kpc}) = 2.37  \left(2^{1/(3-\gamma)}-1\right)M_{{\rm g},11}^{0.14}.
\label{rg}
\end{equation}

\begin{figure}
\centering
\includegraphics[width=8.5cm]{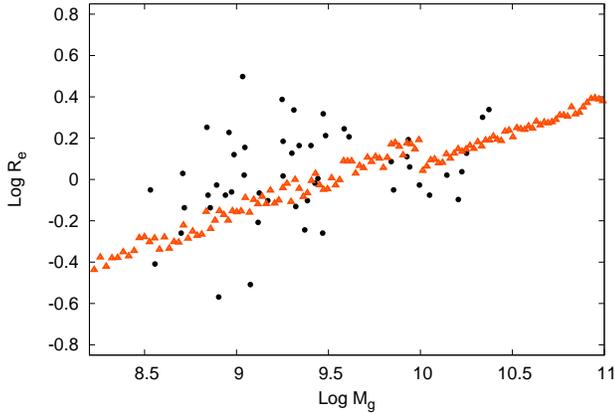}
\caption{Effective radius as a function of the galaxy mass. Black filled circles are data given in LKB12 
while the triangles represent the $R_{\rm e}$ estimated with Equation \ref{effec}.}
\label{MRe}
\end{figure}

In Figure \ref{MRe} the above $R_{\rm e}-M_{\rm g}$ curve is drawn together with observed data taken from LBK12.

On the other hand, to give an estimate of the total radius of the galaxy we 
developed the following relation:

\begin{equation}
	R({\rm kpc}) = 31.62R_{\rm g}{M_{{\rm g},11}}^{1/6} ,
	\label{rtot}
\end{equation}

which allows us to obtain total radii for our galaxy models going from few $\mathrm{kpc}$ for dwarf galaxies to several $\mathrm{kpc}$ for giant ellipticals, 
and give us an estimate of the maximum distance from the galactic centre allowed as initial position for the clusters.

As example, for a galaxy mass $M=10^{12}$M$_\odot$ we obtain a radius $R\simeq 65\-\mathrm{kpc}$ , that is a reasonable value for such galaxies (the supergiant elliptical galaxy M87 has a radius $R\le 100\-\mathrm{kpc}$, comparable to this value).  

The other important parameter that we used to compare our galaxy models with real galaxies is the velocity dispersion, $\sigma_{\rm g}$.

Following LBK12, to evaluate $\sigma_{\rm g}$ we used the formula given in \cite{cappe}:
\begin{equation}
	\sigma_{\rm g}^2=\frac{GM_{\rm g}}{5f_\Omega R_{\rm g}},
	\label{sif}
\end{equation} 
where $f_\Omega=\Omega_{\rm b}/\Omega_{\rm m}$ is the baryonic mass fraction assumed $f_\Omega=0.16$ as in LBK12.

\begin{figure}
\centering
\includegraphics[width=8.5cm]{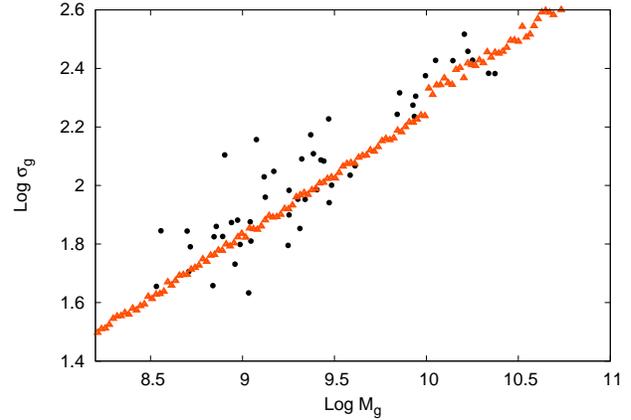}
\caption{Modeled velocity dispersion (triangles) compared with observations (LBK12, filled circles).}
\label{Msig}
\end{figure}

Again, for any given galaxy mass, we selected randomly the parameter $\gamma$ in the ranges explained above. 
The comparison between $\sigma_{\rm g}$ vs $M_{\rm g}$ in LBK12 with our estimate is shown in Figure \ref{Msig}.

Figures \ref{MRe} and \ref{Msig} convince us that we modeled the hosts sufficiently well to obtain
reliable estimation of df times and, as a consequence, reliable values of the NSC masses.

\begin{figure}
\centering
\includegraphics[width=8.5cm]{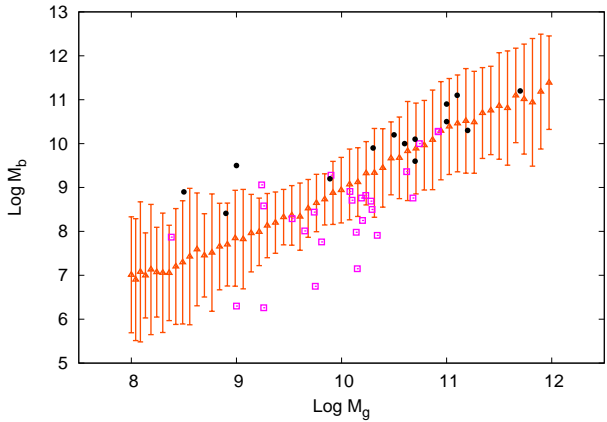}
\caption{Bulge mass in our models (triangles) compared with observed values given in EG12 (squares) and SG13 (filled circles).}
\label{bulGA}
\end{figure}

Several authors had pointed out that the correlation between the bulge and the NSC mass is more dispersed than the $M_{\rm NSC}-M_{\rm g}$ relation \citep{ERWGD}. 
This is also related to the fact that many galaxies are actually bulgeless systems. 
Since the sample of galaxies we used as reference contains also early and late-type spirals, we decided to evaluate the bulge mass for our systems by using the correlation bulge-host given in EG12:
\begin{equation}
{\rm Log}\left(\frac{M_{\rm b}}{10^{9.7}\mathrm{M}_\odot}\right)=(1.23\pm 0.17){\rm Log}\left(\frac{M_{\rm g}}	{10^{9.7}\mathrm{M}_\odot}\right)+(-1.21\pm 0.13).
\label{MbMg}
\end{equation}
This allows us to sample bulges in good agreement with the observed values for spiral galaxies, as it is shown in Figure \ref{bulGA}, and it is still 
coherent to describe the mass of the stellar spheroid in ellipticals and dwarf spheroidal galaxies, where the spheroid mass is indistinguishable from the whole galaxy mass.

Another fundamental ingredient in this framework is the globular cluster system (GCS) total mass. 

A lower limit to the GCS mass can be obtained considering that the ratio between the 
GCS mass and the galaxy mass goes from $\sim 10^{-3}$ for small galaxies ($M_{\rm g} \simeq 10^8$ M$_\odot$), up to $10^{-2}$ for the largest ($M_{\rm g}$ up to $10^{12}$ M$_\odot$). This suggests a weak correlation between the GCS initial mass and the galaxy mass; a good fit formula to this correlation is:

\begin{equation}
	\frac{M_{\rm GCS}(0)}{M_{\rm g}}=6.3\times 10^{-3}M_{{\rm g},11}^{1/6}.
	\label{Mgcs1}
\end{equation}

\cite*{harris14} pointed out that the number of GCs in a given galaxy correlates with the galaxy mass through the relation:
\begin{equation}
{\rm Log} N_{\rm GC} = (1.130\pm 0.098)+(2.810\pm 0.053){\rm Log} (M_{\rm g}/{\rm M}_\odot).
\end{equation}
This relation can be rewritten properly to describe the relation between the galaxy mass and the GCS mass:
\begin{equation}
M_{\rm GCS} = 10^{-8}<m_{\rm GC}>\frac{M_{\rm g}}{\rm{M}_\odot},
	\label{Mgcs2}
\end{equation}
where $<m_{GC}>$ is the mean value of the GC mass in the galaxy. Allowing a mean value of $10^{4}-10^{5}$M$_\odot$ for the GC masses, we recover
a similar expression to Equation \ref{Mgcs1}.

Due to that smaller galaxies host light globulars ($5\times 10^3 - 5\times 10^5$ M$_\odot$), while in heavier galaxies the globulars masses range in the $10^5 - 2\times 10^6$ M$_\odot$ interval \citep{ashmzep}, we set the minimum and maximum value of the GC mass as a function of the galaxy mass:

\begin{align}
	M_{\rm l} = 5\times 10^3\mathrm{M}_\odot \left(4+{\rm Log} M_{{\rm g},11}\right), \label{M1}\\
	M_{\rm u} = 5\times 10^5\mathrm{M}_\odot \left(4+{\rm Log} M_{{\rm g},11}\right). 
\label{M2}
\end{align}

As we will show in Section \ref{stat}, this choice gives mean GC masses
in good agreement with observations.

\section{The merger scenario}
\label{appro}

\subsection{Analytical approach}
\label{ana}
A sufficiently accurate estimate of the NSC mass accumulated to the centre of the galaxy in the merger scenario may be given by means of the following considerations. 
Letting $\Psi(M,r)\mathrm{d}M\mathrm{d}^3\textbf{r}$ be the (infinitesimal) number of GCs with mass in the $[M,M+\mathrm{d}M]$ range and in the volume $\mathrm{d}^3\textbf{r}$ centred at $\textbf{r}$, the total mass of the GCS is:

\begin{equation}
M_{\rm GCS}=\int_V \int_{M_{\rm l}}^{M_{\rm u}} M\Psi(M,r)\mathrm{d}M\mathrm{d}^3\textbf{ r},
\label{MGCS}
\end{equation}

where $M_{\rm l}$ and $M_{\rm u}$ indicate, respectively, the lower and upper value for the GC mass and $V$ is the volume occupied by all the GCs.  
Keeping a sufficient level of generality, we can assume $\Psi(M,r)$ in the form of the product of a function of $M$ and a function of $r$:

\begin{equation}
\Psi(M,r)=\Gamma_0\xi(M)\chi(r),
\end{equation}

where $\Gamma_0$ is a normalization constant given by

\begin{equation}
\Gamma_0=
\begin{cases} 
\displaystyle\frac{(2-s)M_{\rm GCS}}{M_{\rm u}^{2-s}-M_{\rm l}^{2-s}}\displaystyle\frac{1}{N} & s\neq 2, \\
\displaystyle\frac{M_{\rm GCS}}{\ln\left(M_{\rm u}/M_{\rm l}\right)}\displaystyle\frac{1}{N} & s=2,\\
\end{cases}
\end{equation}
with $N=\int_0^R \chi(r) d^3\bf{r}$ the total number of clusters in the galaxy.

A suitable expression for the mass function, $\xi(M)$, is a (truncated) power-law (see for instance \cite{bmgrt}):
\begin{equation}
\xi(M)= M^{-s}.
\label{imf}
\end{equation}

On the other hand, the distribution of radial positions is, in principle, arbitrary.

The simple inversion of Equation \ref{tdf} yields the maximum radius, $r_{\rm max}$, which contains all the clusters with mass $\geq M$ and with initial eccentricity $\leq e$ that have been confined around the galactic centre in a time $\leq t$:

\begin{equation}
r_{\rm max}=R_{\rm g}\left(\frac{t}{A_\gamma}\right)^{0.57}\left(\frac{M}{M_{\rm g}}\right)^{0.38},
\label{RMAX}
\end{equation}
with $A_\gamma=\tau_0(2-\gamma)(1+g(e))$.

Consequently, an estimate of the NSC mass as a result of the accumulation of GCs to the galactic centre, caused by dynamical friction, is:

\begin{equation}
M_{\rm NSC}(t)=\Gamma_0 \int_{M_{\rm l}}^{M_{\rm u}} M^{1-s} N(r_{\rm max}) \mathrm{d}M,
\label{Mnsct}
\end{equation}

with $N(r_{\rm max})$ given by:
\begin{equation}
N (r_{\rm max})=4\pi \int_{0}^{r_{\rm max}} \chi(r)r^2dr.
\label{mgcsmax}
\end{equation}

Let us now consider two different radial distributions for the GC population: a generic power-law distribution, $\chi(r)=Kr^\delta$, and a $\gamma$ model density law (Equation \ref{Dhn}).

In the first case, Equation \ref{mgcsmax} reduces to:
\begin{equation}
N(r)=
\begin{cases}
\displaystyle N \left(\frac{r}{R}\right)^{\delta+3} & r\le R,\\
\displaystyle N    & r> R,\\
\end{cases}
\end{equation}
with $N$ the total number of clusters in the galaxy.

By substitution of this relation into Equation \ref{Mnsct} we obtain:

\begin{equation}
M_{\rm NSC}(t)= \Gamma C^n\int_{M_{\rm l}}^{M_{\rm u}}M^{1-s+0.38n}\mathrm{d}M,
\label{MNSC}
\end{equation}
with $\Gamma=\Gamma_0 N$.

where $n=\delta+3$ and $C$ is a function of $M_{\rm g},R_{\rm g},\gamma,e$ and $t$ 
whose explicit expression is:
\begin{equation} \nonumber
C=\left(t/A_\gamma\right)^{0.57}\left(1/M_{\rm g}\right)^{0.38}R_{\rm g}.
\end{equation}

After integration, Equation \ref{MNSC} yields
\begin{equation}
M_{\rm NSC}(t)=\Gamma C^{n}\frac{M_{\rm u}^{2-s+0.38n}-M_{\rm l}^{2-s+0.38n}}{2-s+0.38n}.
\label{tAN}
\end{equation}

Looking at Equation \ref{tAN}, it is evident now the dependence of the NSC mass from the $n$ parameter, i.e. from the steepness of the 
density profile.

The more general case in which we consider a $\gamma$ density law, instead, will be discussed in the Appendix.

\subsection{Results of the analytical approach}

Allowing $\delta$ to vary in Equation \ref{tAN} we estimate the mass of the NSCs for different values of the slope of the GCs mass function $s$ at varying the galaxy mass in the range $[10^8-10^{12}]$M$_\odot$.

Equation \ref{tAN} allows us to see how the NSC mass increase as a function of time.

\begin{figure}
\includegraphics[width=8.5cm]{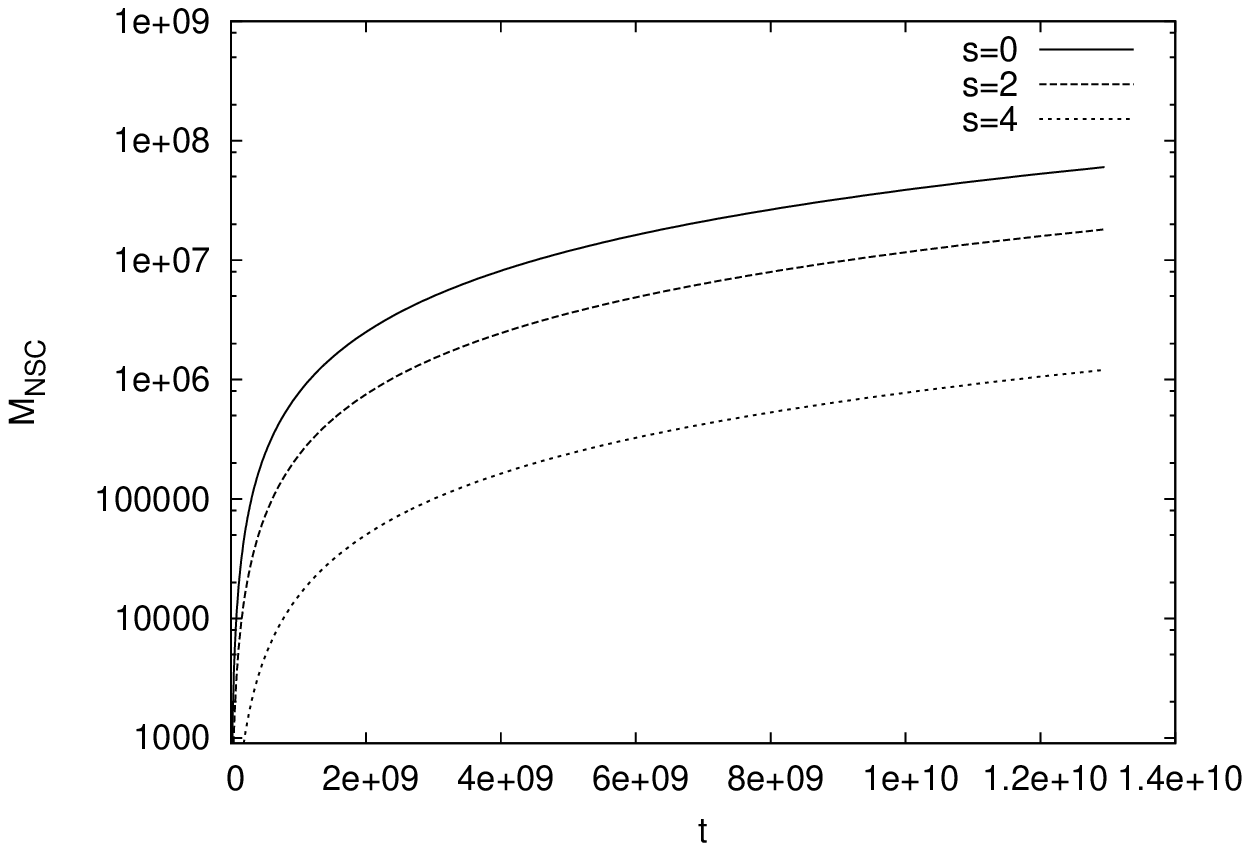}\\
\includegraphics[width=8.5cm]{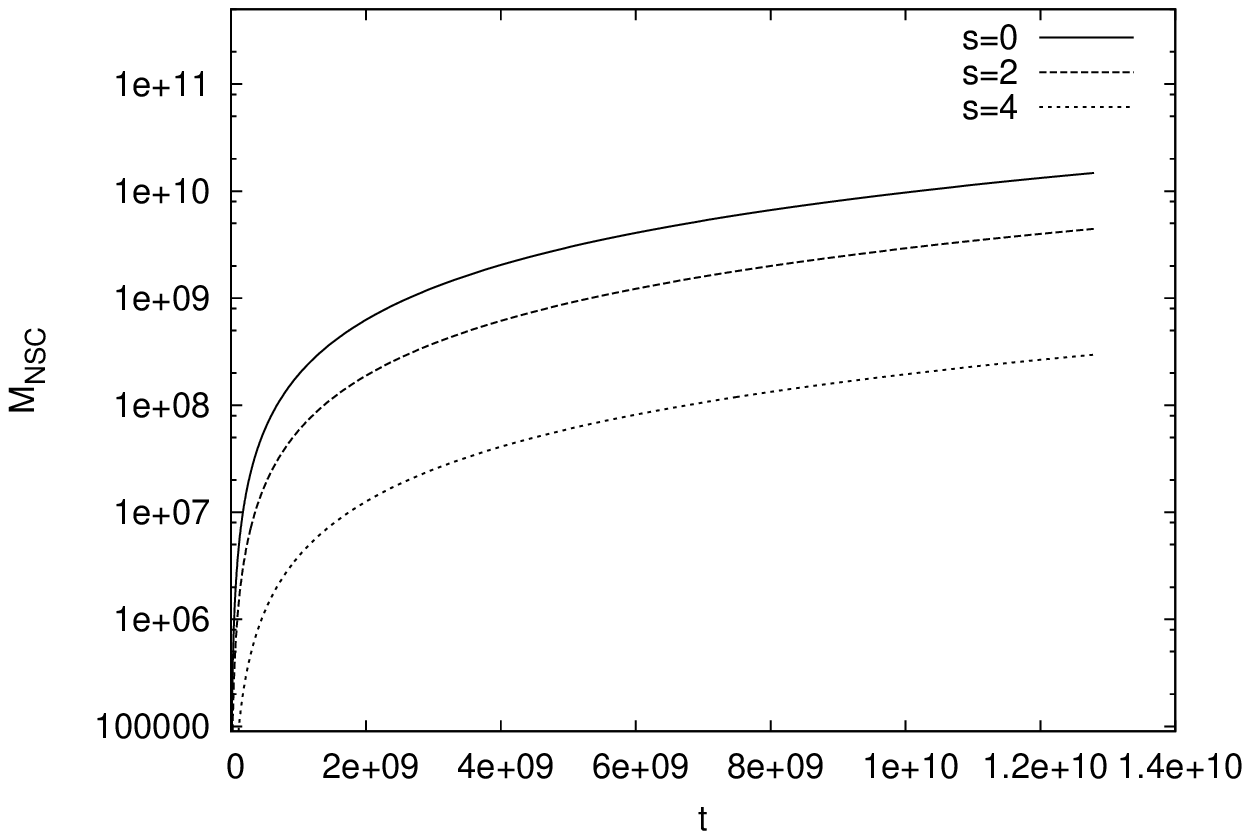}
\caption{NSC mass growth for a galaxy with $M_{\rm g}=10^8$M$_\odot$ (upper panel), and $M_{\rm g}=10^{12}$M$_\odot$ (bottom panel).}
\label{growth}
\end{figure}

Figure \ref{growth} shows the NSC growth as a function of time in the case $\delta=0$ for two extreme values of the galaxy mass ($10^{8}$ and $10^{12}$ M$_\odot$) and three values of $s$ considering the mass function in Equation \ref{imf}, i.e. $s=(0,2,4)$ with $M_{\rm l}$ and $M_{\rm u}$ as defined by Equation \ref{M2}.

The figure shows that the NSC mass increases rapidly in an early phase ($t \le 1$ Gyr) to slow down later its growth. The slower increase in the case of larger values of $s$ depends on the smaller fraction of heavy, and fast decaying, GCs for steeper mass functions.

\begin{figure}
\centering
\includegraphics[width=8cm]{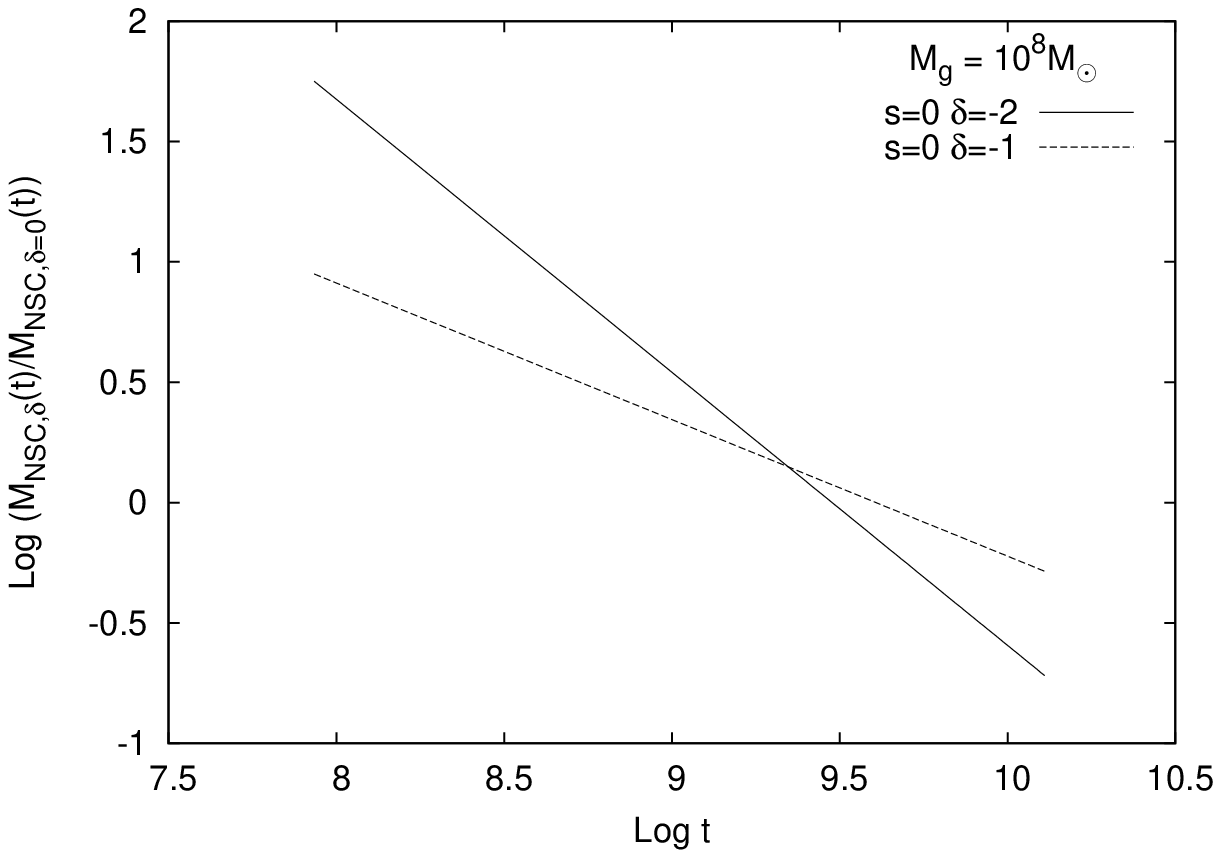}
\includegraphics[width=8cm]{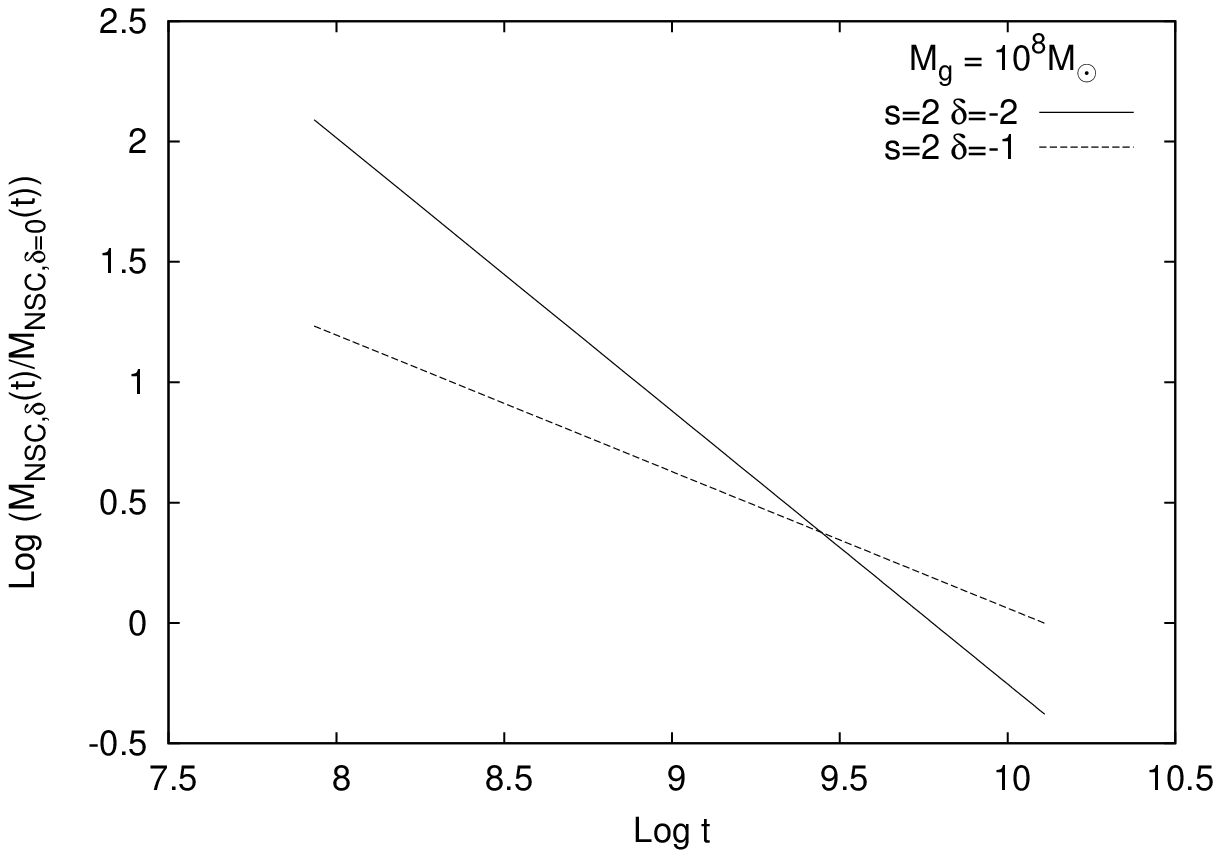}
\includegraphics[width=8cm]{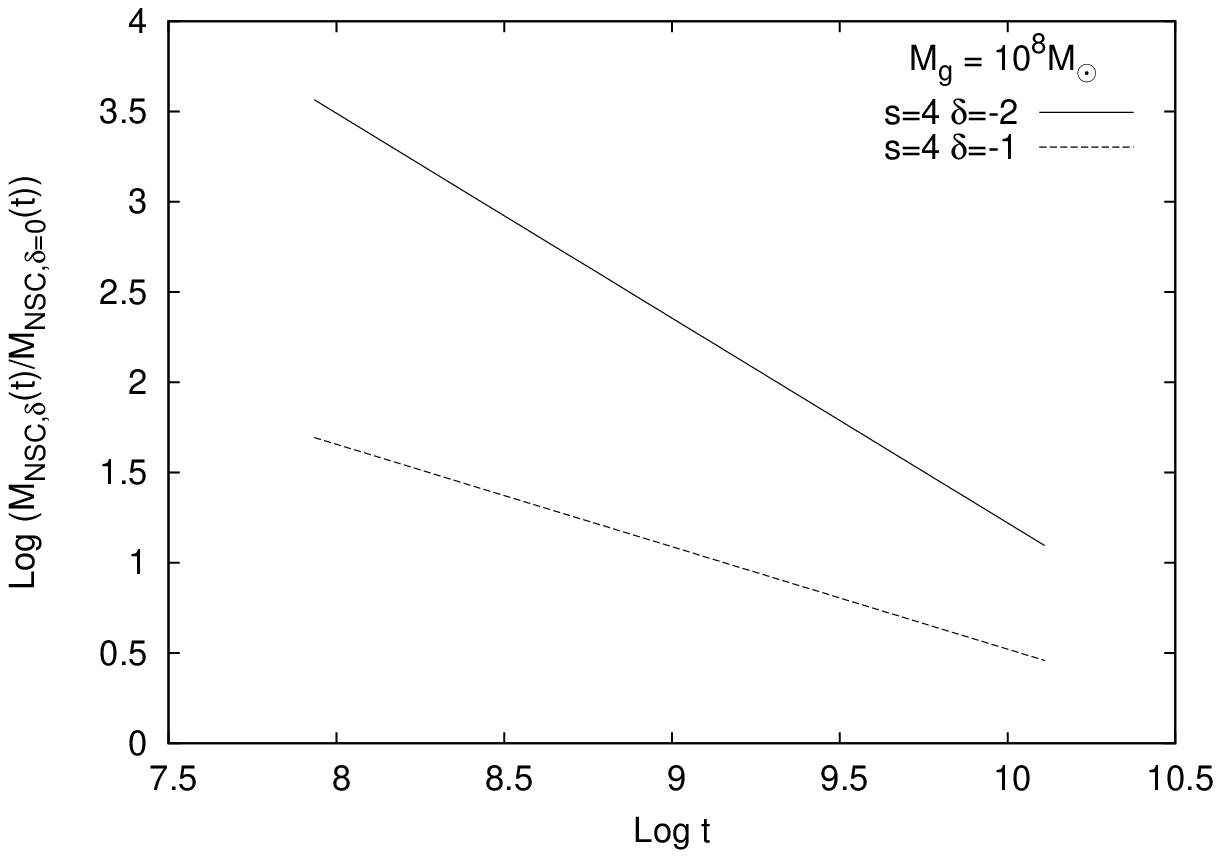}
\caption{The ratio between the mass of NSC for a galaxy whose mass is $M_g=10^8$M$_\odot$ evaluate with $\delta=-2$ (straight line) and $\delta=-1$ (dotted line) and $M_{NSC}$ obtained lettin $\delta=0$. From top to bottom, we set $s=0,\- 2,\- 4$.}
\label{growthcmp1}
\end{figure}

Using the case $s=2$, $\delta=0$ as reference, we show in Figure \ref{growthcmp1} and \ref{growthcmp1} the ratio between the NSC mass evaluated letting $\delta=-1$, $-2$ and that obtained with $\delta=0$ at fixed $s$ and $M_g$.

\begin{figure}
\centering
\includegraphics[width=8cm]{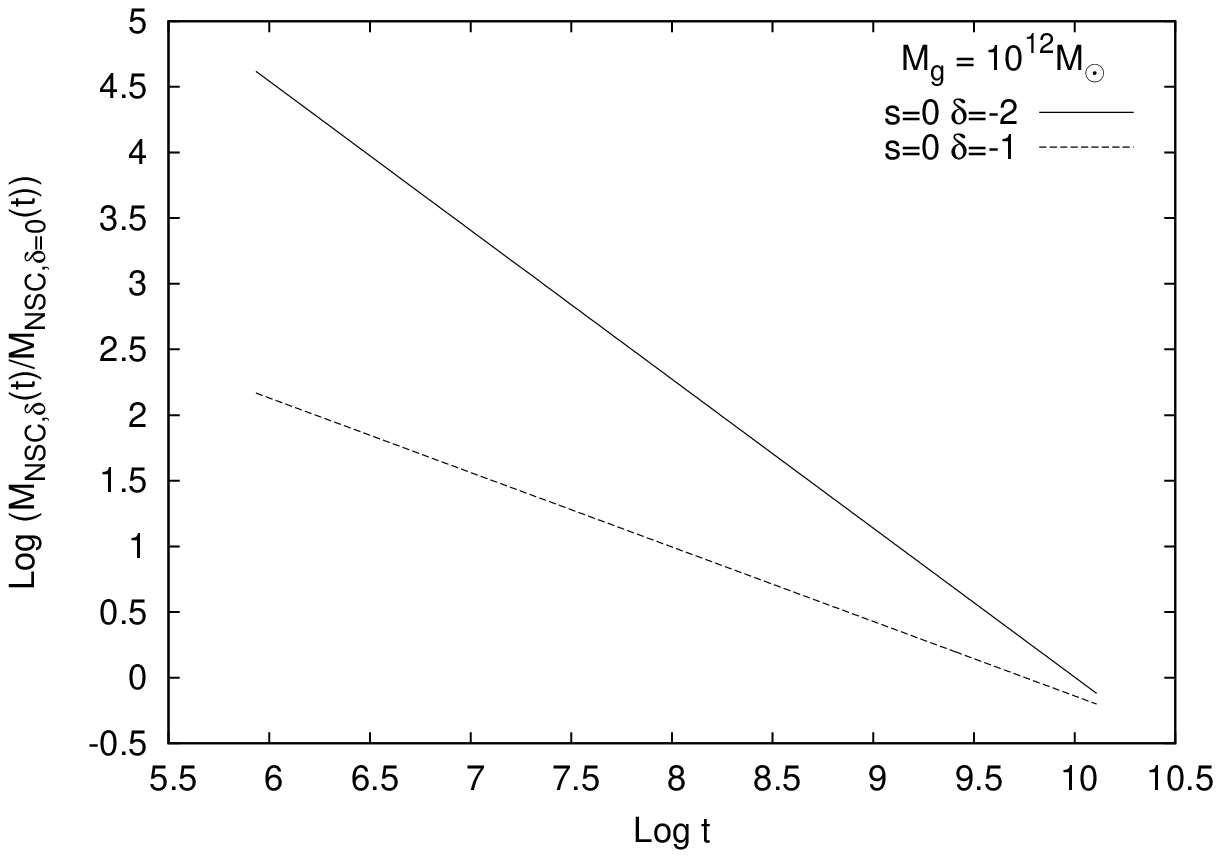}
\includegraphics[width=8cm]{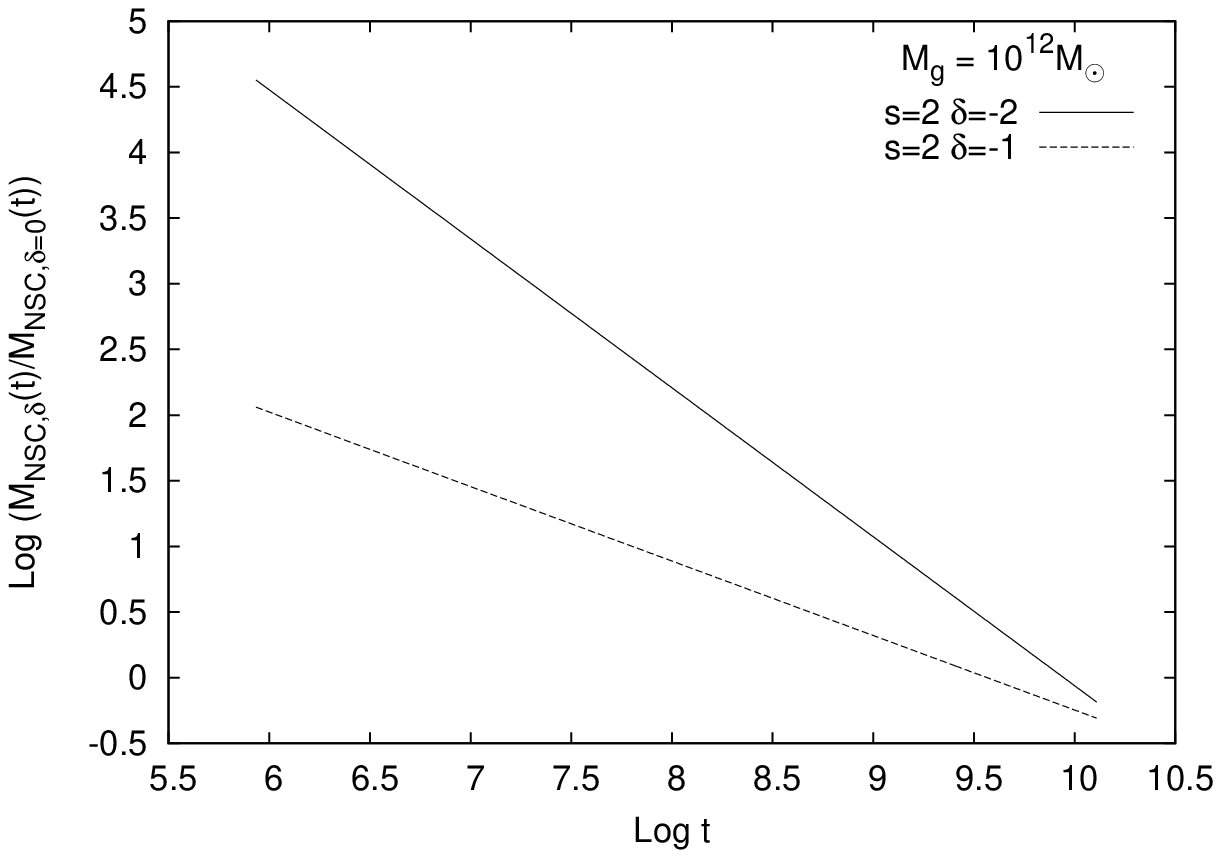}
\includegraphics[width=8cm]{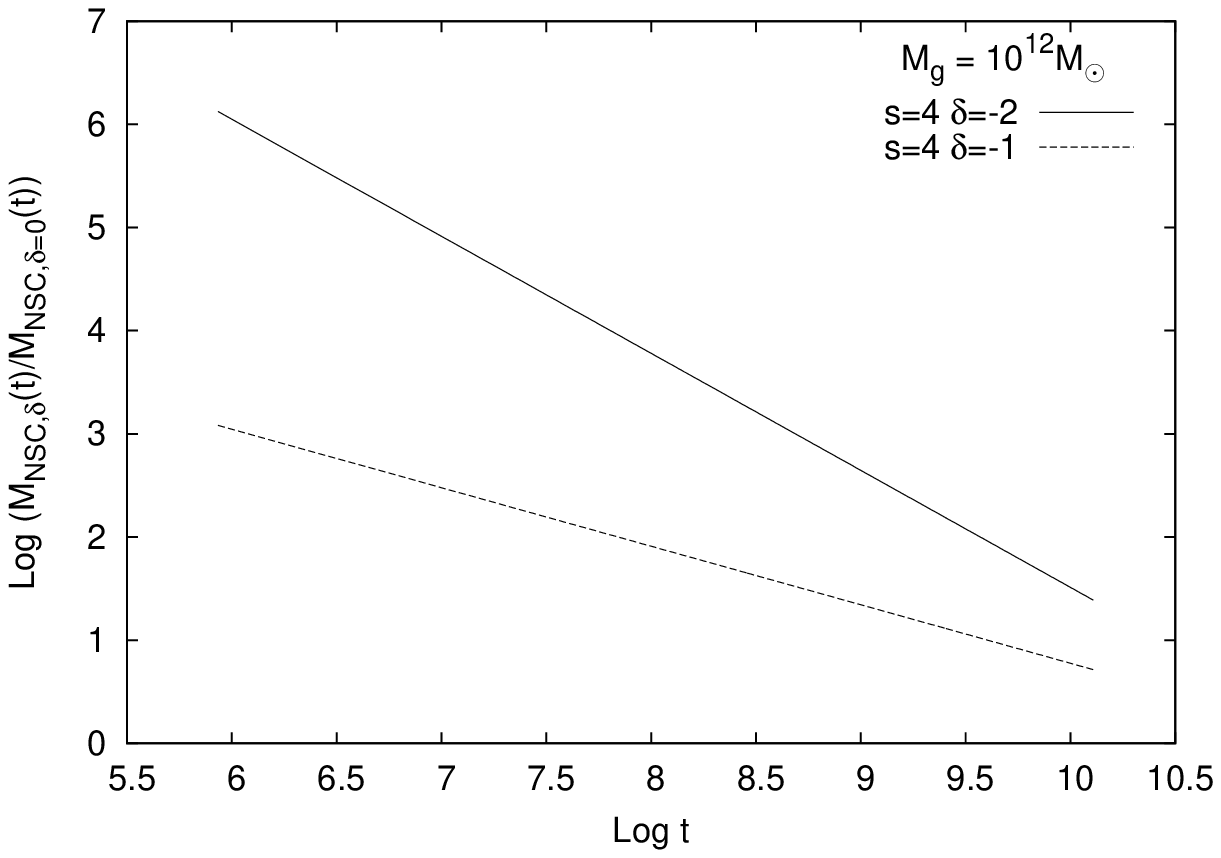}
\caption{The same as in Figure \ref{growthcmp1}, but for a galaxy with $M_g=10^{12}$M$_\odot$.}
\label{growthcmp2}
\end{figure}

Considering the case $M_g=10^8$M$_\odot$, it is evident that in an early phase ($t<10^8yr$) the smaller $\delta$ the faster the NSC mass growth; however
as the time increase is evident that the final mass of the cluster is slightly small with respect to the reference case $\delta=0$ if $s=0,\- 2$, while 
considering $s=4$ the smaller the $\delta$ the greater the final mass of the NSC.

Considering instead more massive galaxies ($M_g=10^{12}$M$_\odot$) we found that the smaller the $\delta$ the greater the final NSC mass.

\begin{figure}
\centering
\includegraphics[width=8cm]{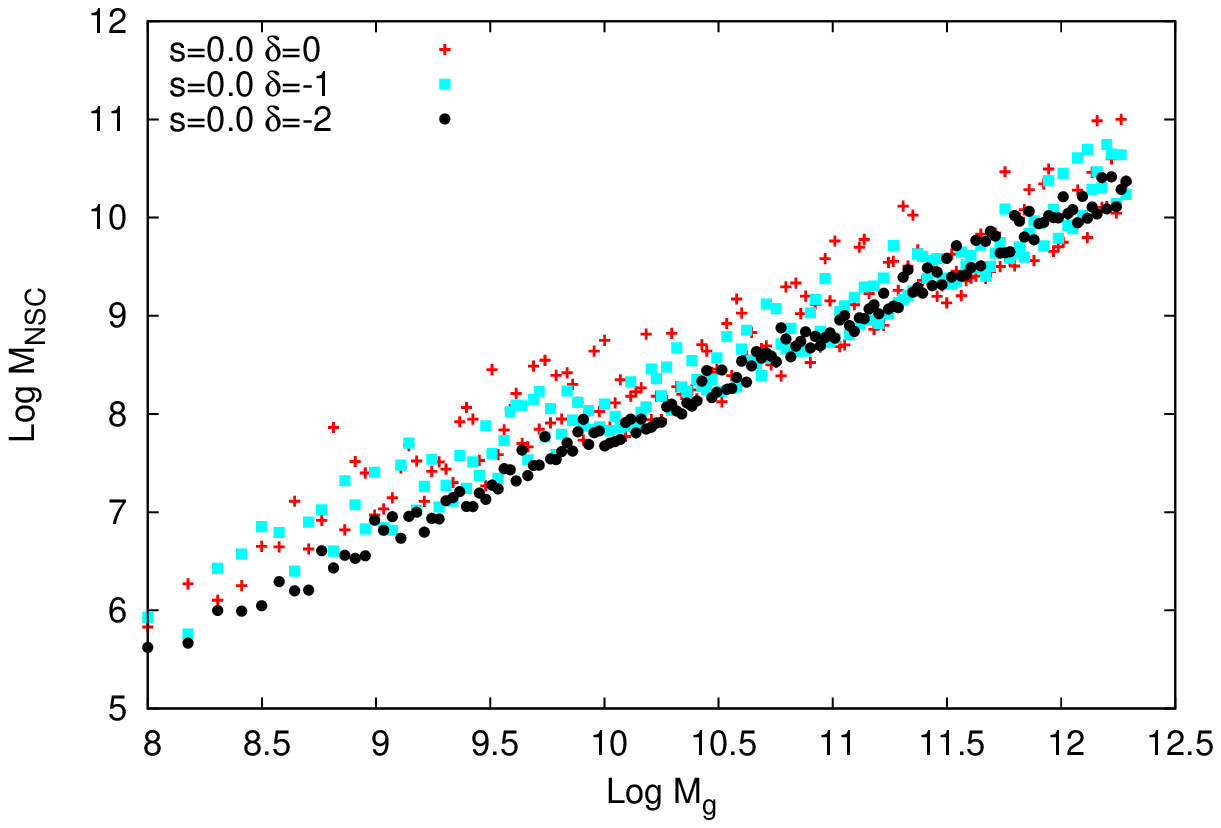}
\includegraphics[width=8cm]{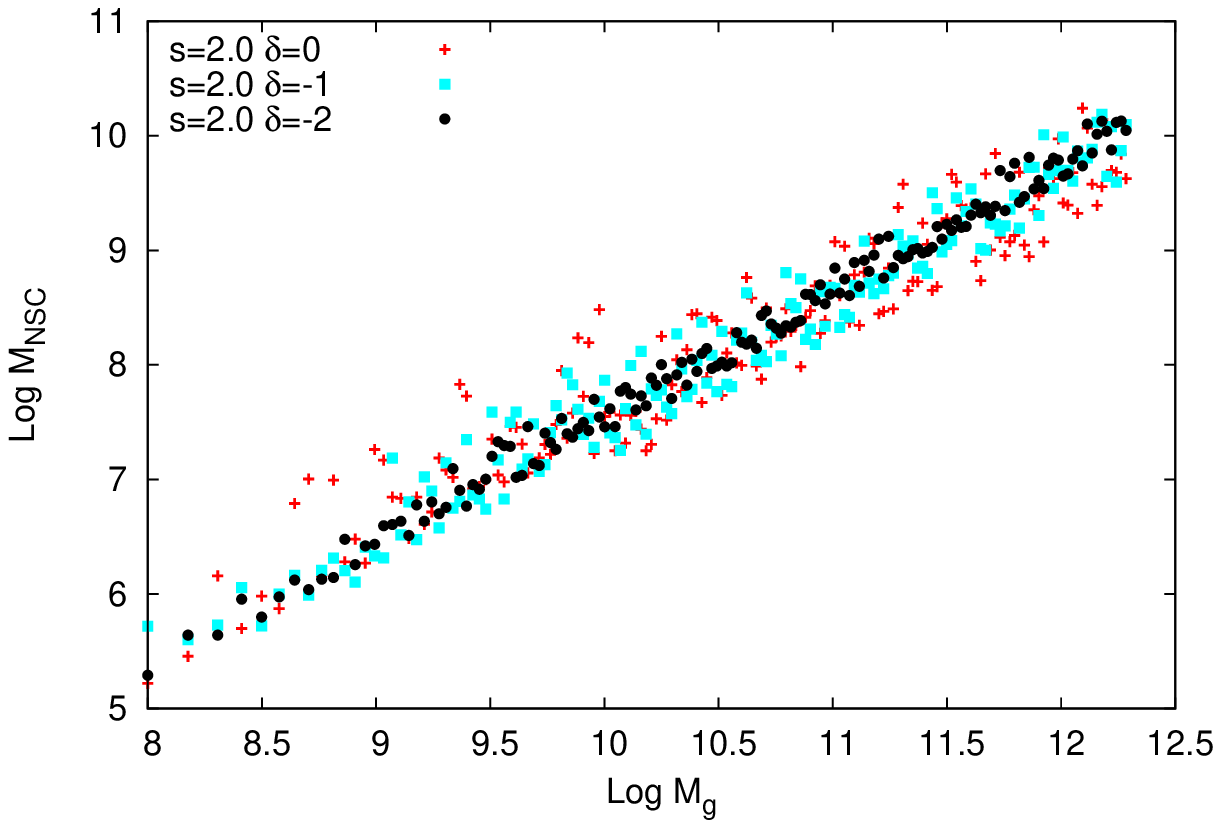}
\includegraphics[width=8cm]{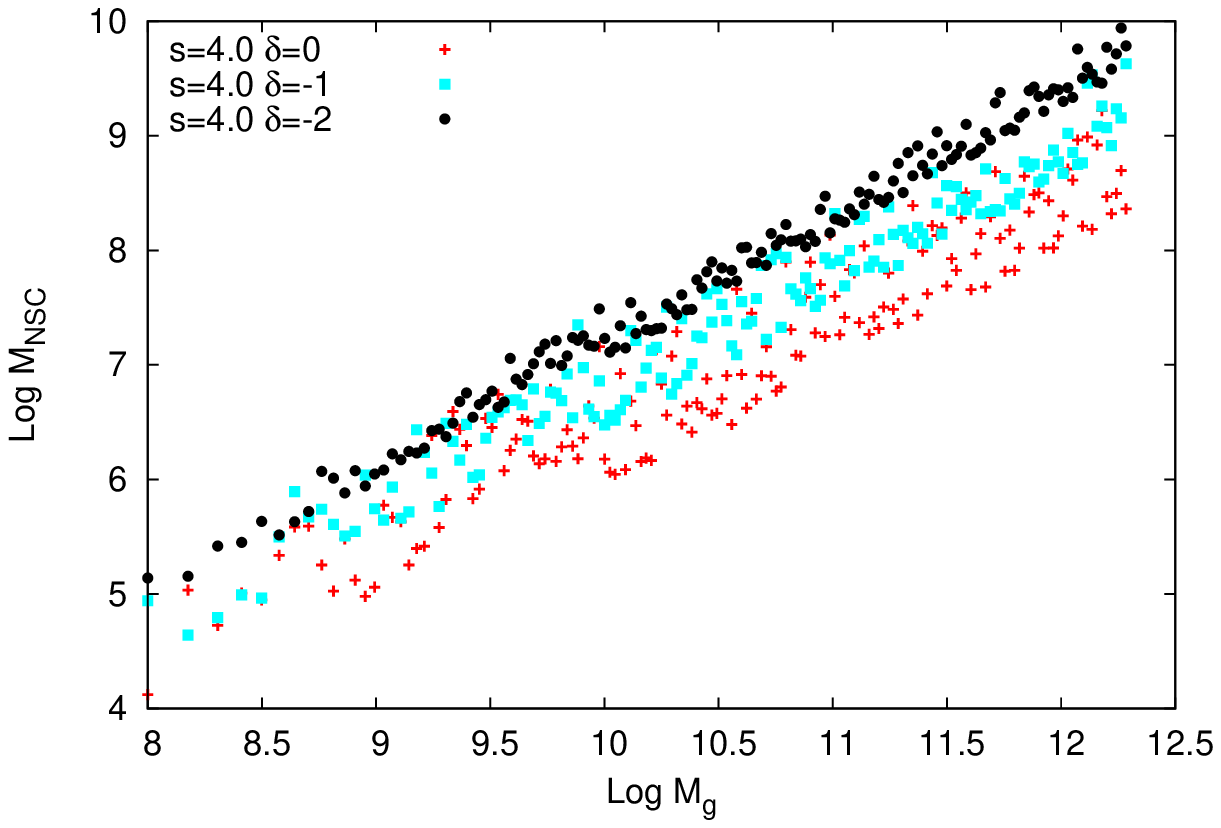}
\caption{Masses of NSCs as a function of the host mass for different values of $s$ and $\delta$. From top to bottom, we set $s=0,\- 2,\- 4$. In each
panel, are shown results for $\delta=0,\- -1,\- -2$.}
\label{MnscCmp}
\end{figure}

Figure \ref{MnscCmp} shows the Mass of NSCs as a function of the host mass for $\delta=0,-1,-2$ and $s=0,2,4$. At any fixed value of $s$, there 
is not a significant difference between NSC masses estimate with different values of $\delta$.

\begin{figure}
\includegraphics[width=8.5cm]{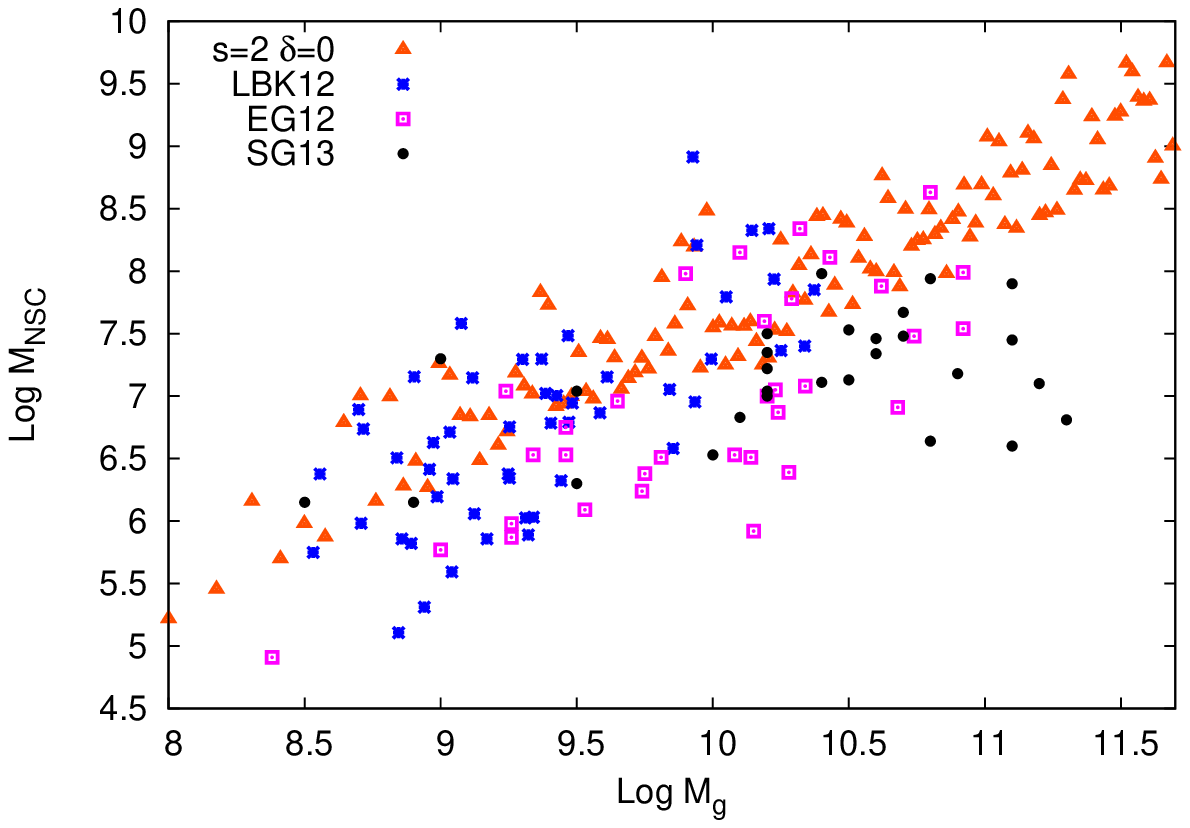}\\
\includegraphics[width=8.5cm]{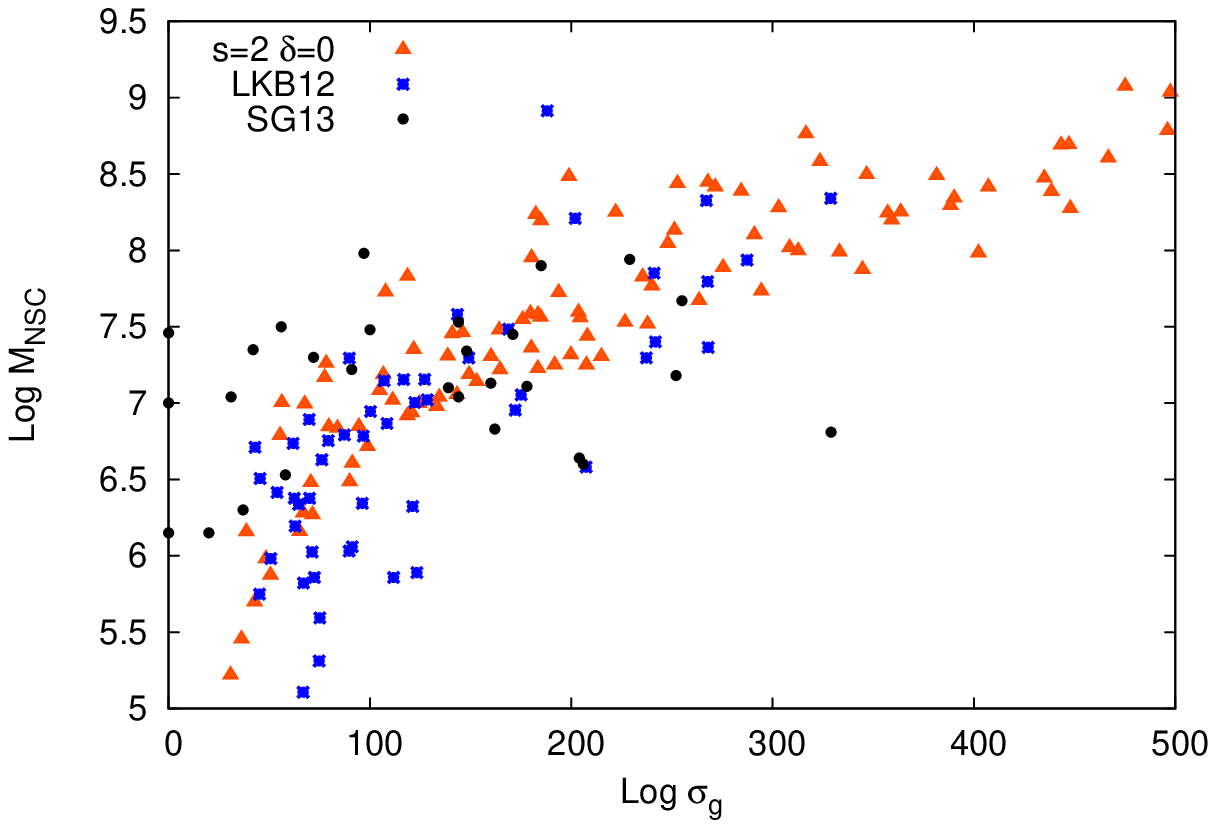}\\
\includegraphics[width=8.5cm]{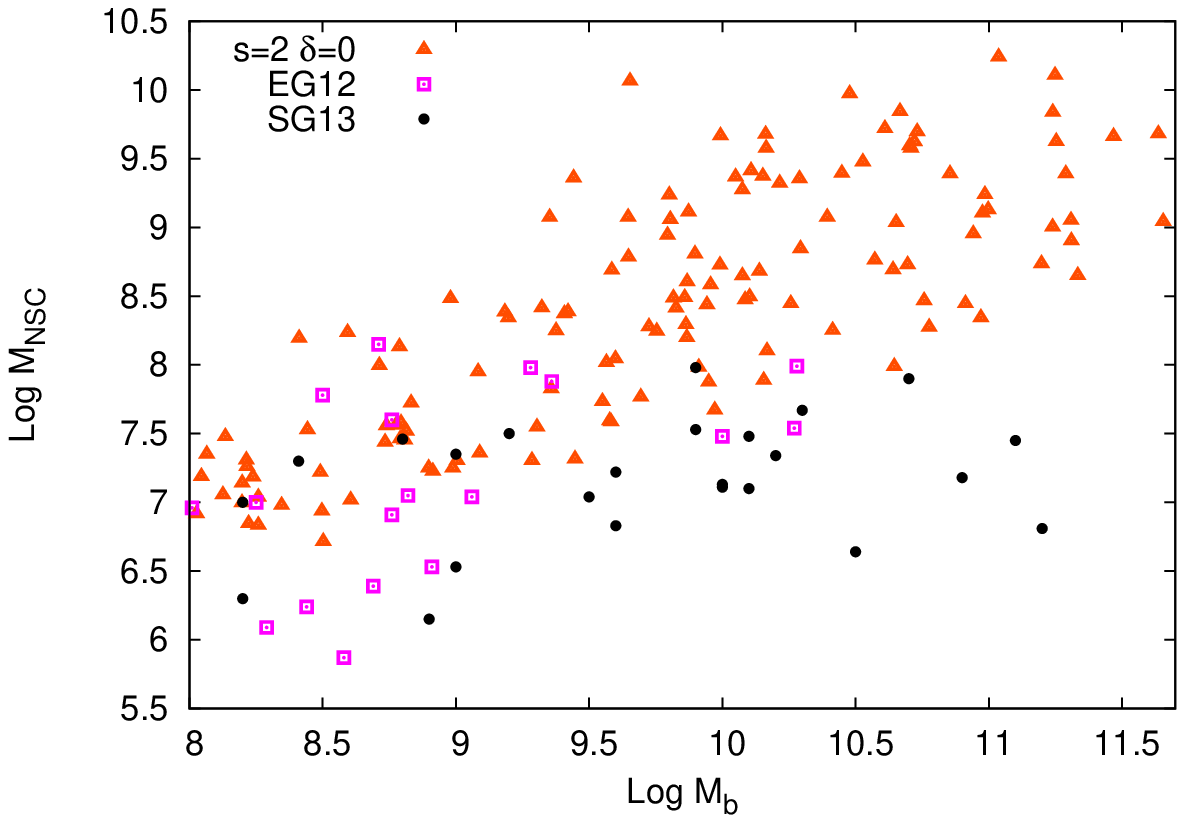}\\
\caption{Theoretical NSCs masses vs. the host properties (triangles) estimated by using Equation \ref{tAN} with $s=2$, $\delta=0$; compared with NSC masses from data in LKB12 (stars), EG12 (squares) and SG13 (filled circles). From top to bottom, the various panels give the correlation with hosts masses, velocity dispersion and bulge mass.}
\label{NSCM}
\end{figure}

\begin{figure}
\includegraphics[width=8.5cm]{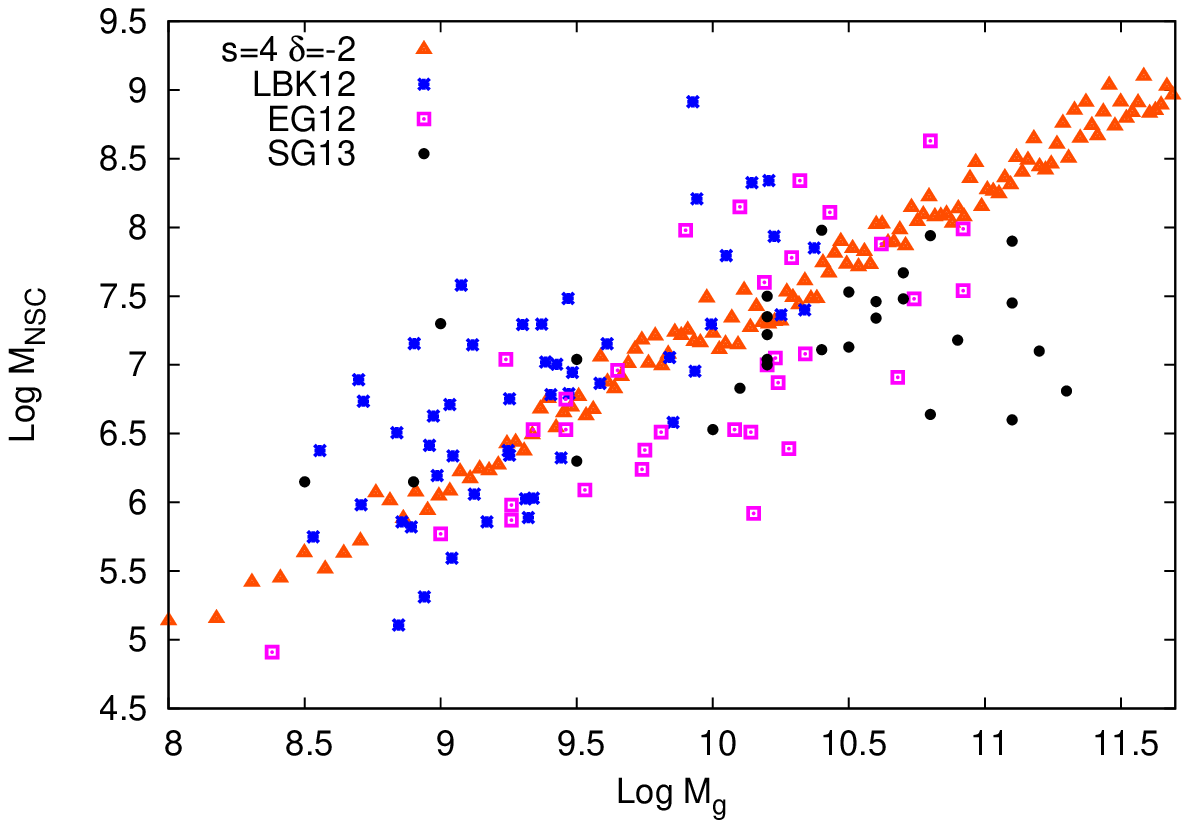}\\
\includegraphics[width=8.5cm]{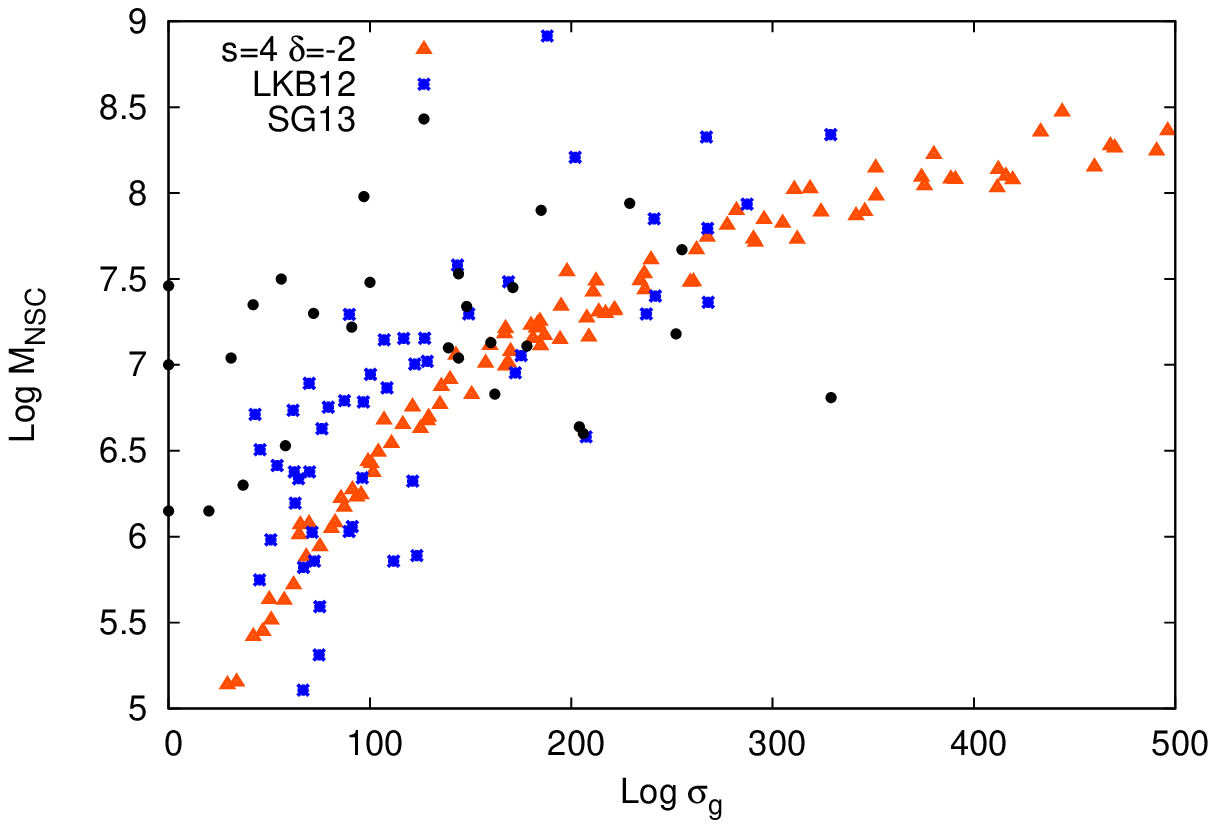}\\
\includegraphics[width=8.5cm]{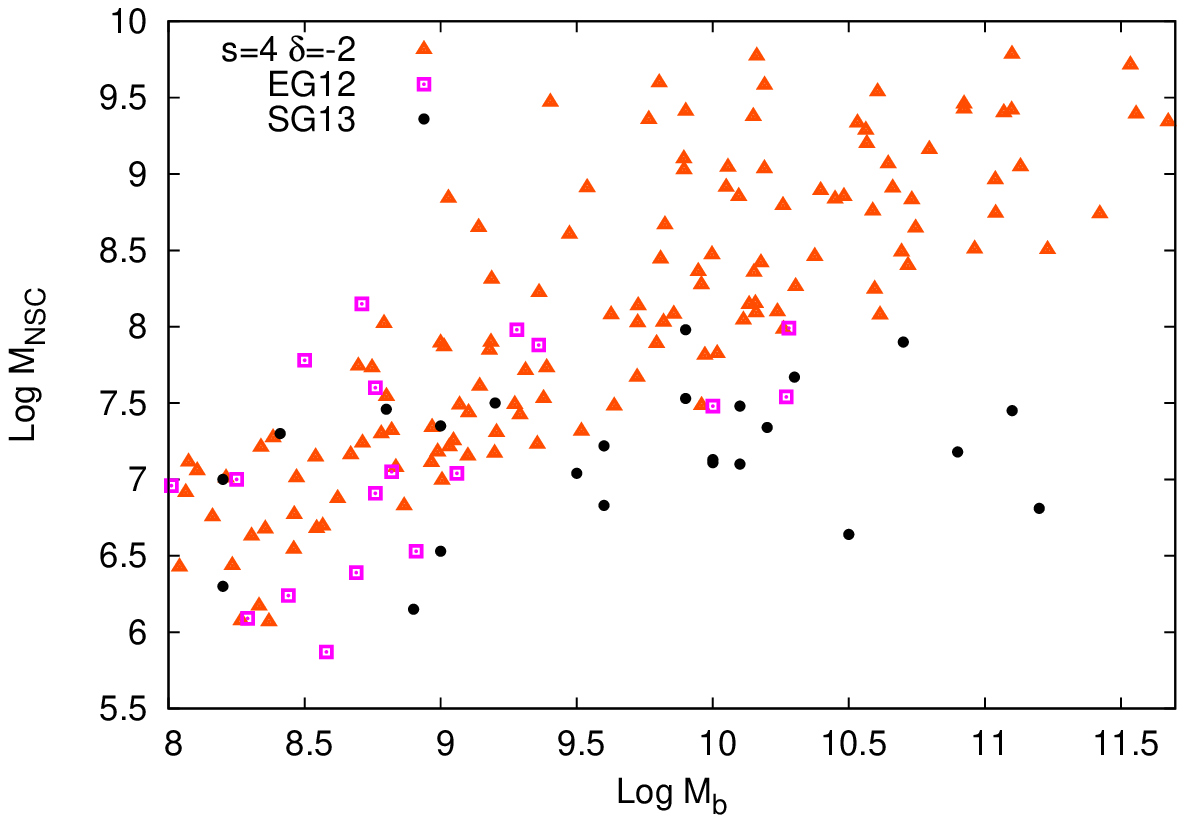}\\
\caption{The same as in Figure \ref{NSCM}, but in the case $s=4$, $\delta=-2$.}
\label{NSCM2}
\end{figure}

In Figure \ref{NSCM} we compare the theoretical NSC mass (that evaluated at $t = 13$ Gyr)  with the observational values from EG12 and LBK12. The best agreement is achieved choosing $s=2,\- \delta=0$, as it will be more deeply discussed in Section \ref{laws}. Moreover, we found that a good 
correlation is achieved in the case $s=4,\- \delta=-2$, as it is shown in Figure \ref{NSCM2}, but this extreme case in which both the density profile
and the mass function are very steeps, is really unlikely to reproduce real galaxies. 
For this reason, we limited the analysis to the case $s=2,\- \delta=0$. In the following, we refer to the model $s=2$ as the combination ($s=2,\- \delta=0$).

In this context, it is relevant noting that the mass distribution of young luminous clusters (often referred to as YLCs) in many galaxies is a power-law with a spectral index ranging between $s=1.8$ and $s=2$ \citep{Whitmore}. Assuming such power-law as initial mass function for a GCS, \cite{bmgrt} showed that the evolution of such a system in a model of our galaxy, leads to a final mass function for the GCs in agreement
with the actual mass function of the Milky Way GCS, giving us some confirmation about the choices made to model GCSs.

\subsection{Statistical approach}
\label{stat}

Beside the 'analytical' method illustrated above to estimate the NSCs masses, we investigate the infall scenario also from a 'statistical' point of view, in order to obtain information on the radial distribution of GCs in the host galaxy, the number of globulars centrally decayed and that of GC survivors.

The idea behind this statistical approach is sampling the initial GCS of a given galaxy, and evaluate how many GCs sink toward the galactic centre within a Hubble time. A statistical estimate of the expected NSC mass is thus obtained by making $N_s$ realizations of the GCS of a galaxy, in order to give constraints to the error. Each galaxy was modeled as explained in Section \ref{model}, while, for each cluster, we sampled its initial radial position, $r_0$, and orbital eccentricity, $e$, from a random, flat distribution.

From this GCS sampling we infer two relevant parameters to compare with observations.
The first is the mean value of the GC mass for any given host mass, that 
can be compared with data given in LKB12; the second parameter is
the number of survived clusters, which goes from few ($<10$) GCs  
for small galaxies ($M_{\rm g}\sim 10^8$M$_\odot$), to few hundreds in 
intermediate mass galaxies, and up to $10^4$ in giant ellipticals.

The relatively strong dependence of the df braking time on the individual GC mass (see Equation \ref{tdf}) deserves a careful treatment in the GC sampling, as explained in detail here below.

\subsection{The statistical GCS modelization}
\label{statmod}
As mentioned above, in this section we give estimate of the expected NSC mass for a given galaxy mass, sampling the whole GCS of the galaxy and looking at which clusters can sink toward the galactic centre within a Hubble time. Since initial position, eccentricity of the orbits and mass of each cluster are fundamental parameters in the evaluation of the decay time (see Equation \ref{tdf}), we vary the sampling method for the GCS, changing spatial distribution and mass function of the clusters as explained in the following.

\subsubsection*{Flat radial density and mass power-law sampling (PLS)}

This model (referred to as PLS model) is characterized by a flat spatial distribution of GCs within the radial  range $[0-R]$, with $R$ the maximum distance defined in Equation \ref{rtot}; their eccentricities are sampled randomly between $0$ and $1$. GC  masses are distributed according to a power-law distribution, $\mathrm{d}N \propto M^{-s}\mathrm{d}M$. PLS is, actually, the 'statistical version' of the analytical treatment (see Section \ref{ana}).
When $s=0$ (flat mass distribution) we refer to as the random sampling model, called RND.

\subsubsection*{Flat radial density and mass Gaussian sampling (GSS)} 
In the GSS model, the spatial location of GCs is the same as above, while the mass sampling is made by means of a gaussian generator, given a mean value $M_{mean}/\mathrm{M}_\odot=10^5\left[4+{\rm Log} M_{{\rm g},11}\right]$,  
and a fixed dispersion, $\sigma=0.25$.

To exclude unrealistic, too light or too massive objects in the mass distribution, we truncated the gaussian at a low mass $M_{\rm l} = 5\times 10^3$M$_\odot$ and at a high mass $M_{\rm u} = 2\times 10^6$M$_\odot$.

\subsubsection*{$\rho_{\gamma}$ radial density sampling (RHO).}
There is no compelling evidence that the GCs and stars of the parent galaxy followed, initially, different density profiles, so we found worth examining the case where the initial GCS density profile is the same $\gamma-$ profile of the parent galaxy.

\subsection{Results of the statistical approach}
\label{res}

The quality of our GC sampling can be tested by a comparison of the GCs mean masses obtained, $M_{\rm GC}$, with data given in LKB12.

\begin{figure}
\centering
\includegraphics[width=8cm]{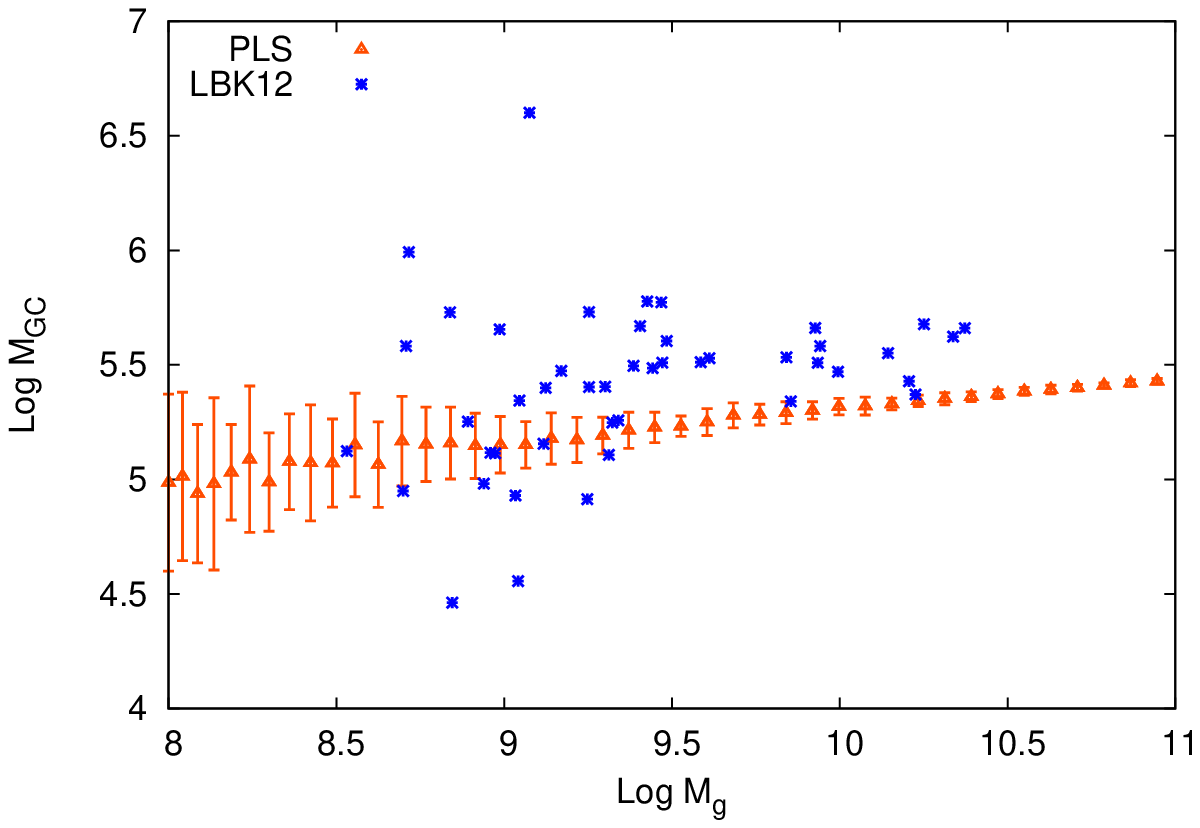}\\
\includegraphics[width=8cm]{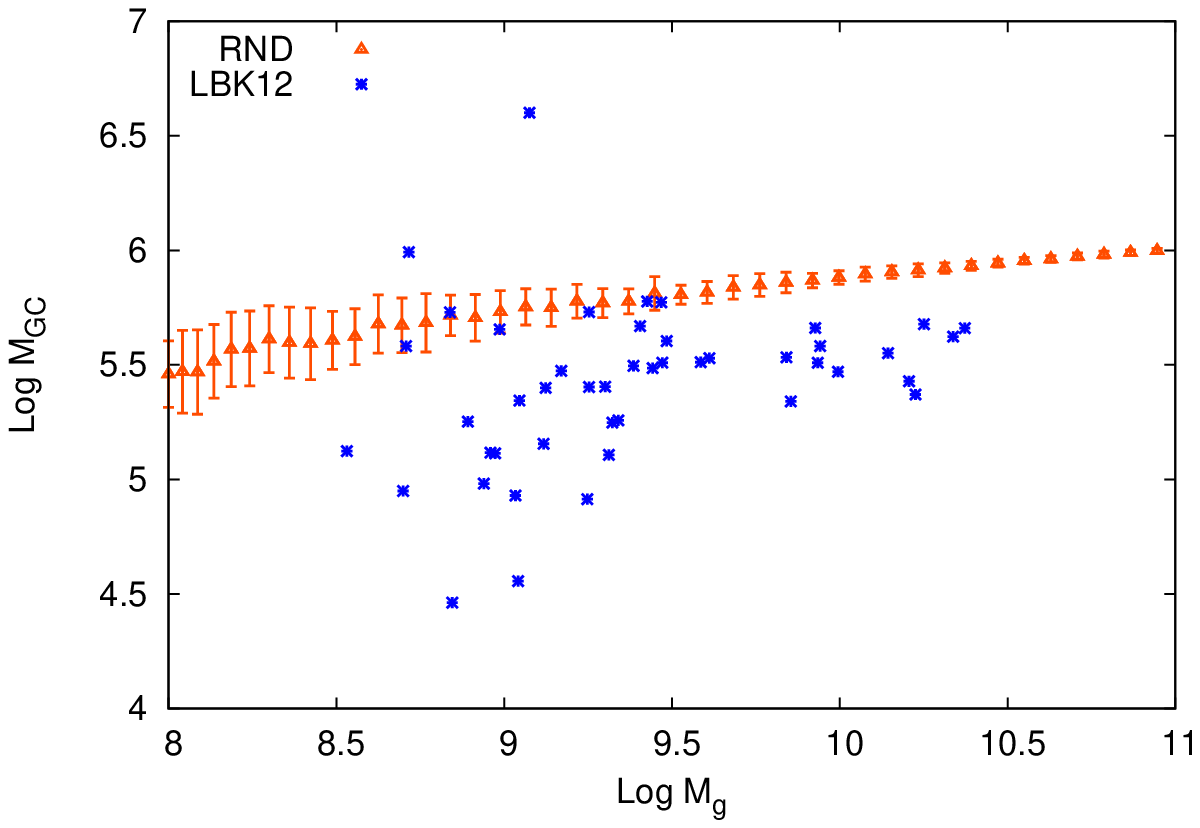}
\includegraphics[width=8cm]{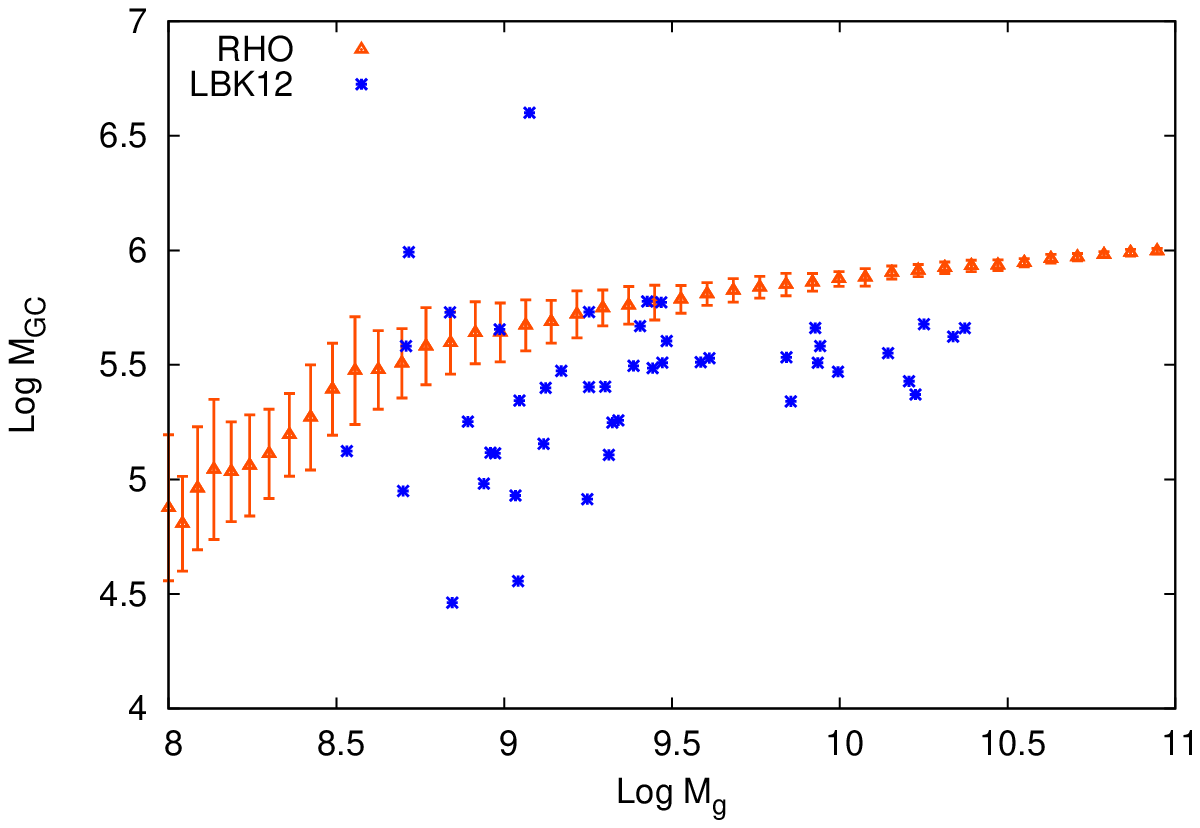}
\includegraphics[width=8cm]{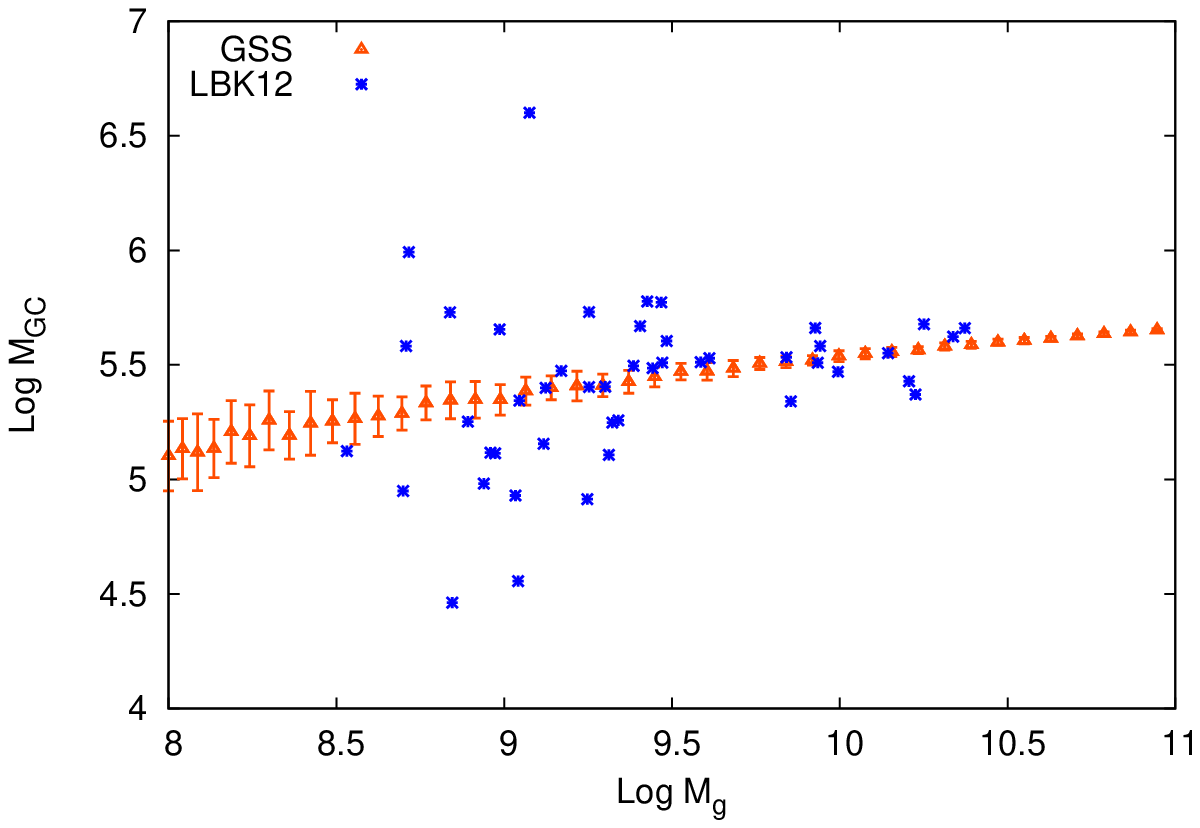}
\caption{From top to bottom, mean GC mass in models PLS, RND, RHO and GSS as a function of the galaxy mass.
Triangles represent our theoretical data while crosses represent the observations.}
\label{mgc}
\end{figure}

Looking at Figures \ref{mgc}, we see that GSS and RHO models give a decent agreement with observations in the whole range of masses covered by the data available; on the other hand, the RND model seems to overestimate the mean GC mass, while the PLS model gives an underestimate for host masses above $2\times 10^9\rm M_\odot$.

\begin{figure}
\centering
\includegraphics[width=8.5cm]{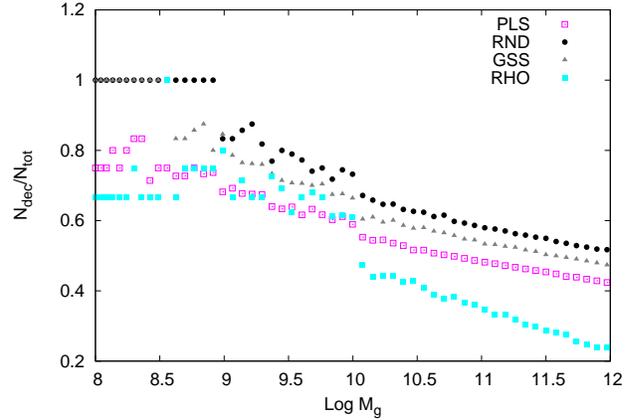}
\caption{Number of decayed clusters over the total number of clusters within a Hubble time as a function of the host mass for each model considered.}
\label{srv}
\end{figure}

\begin{figure}
\centering
\includegraphics[width=8.5cm]{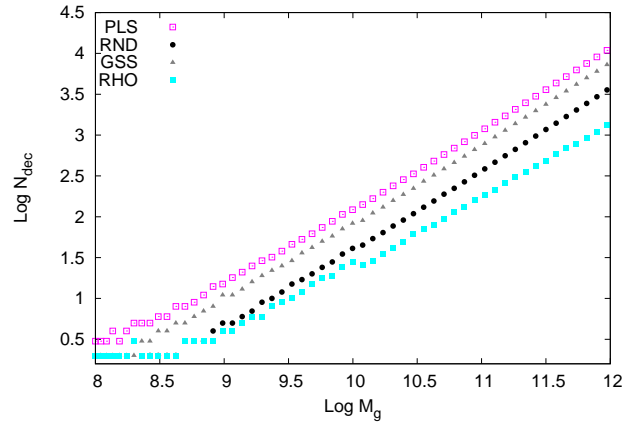}
\caption{Number of decayed clusters within a Hubble time as a function of the host mass for each model considered.}
\label{srv2}
\end{figure}

Another relevant quantity obtainable with this approach is the number of 'survived" clusters, which can be compared with the actually observed clusters. In Figure \ref{srv} the fraction of decayed clusters as a function of the hosting galaxy mass is shown. Moreover, Figure \ref{srv2} shows also the number of decayed clusters.

It should not surprise that for sufficiently massive galaxies ($M_{\rm g}>10^{10}$M$_\odot$) the number of survived
GCs can exceed $10^4$. Many massive galaxies actually host such large populations of clusters. As example, the giant elliptical galaxy M87 (also known as Virgo A, with a mass $\sim 2\times 10^{12}$ M$_\odot$ \citep*{laughHar} hosts about $13,000$ clusters, in agreement with our prediction.

On the other hand, the small number of GC expected in galaxies with $M_g\sim 10^8$M$_\odot$, could provide a possible explanation for the lack of nucleated region in dwarf spheroidal galaxies in the Virgo cluster \citep{VdB}. In fact in such small galaxies, only few clusters are expected to form, this would imply that the formation of a NSC, at least in the framework of the dry-merger scenario, depends strongly on the initial conditions of each cluster of the galaxy. Fluctuations in this low number statistics, together with the low mass of GCS in these environments (and so long dynamical friction times) make the probability of finding NSCs in the center of these small hosts very low.

\begin{figure}
\centering
\includegraphics[width=8cm]{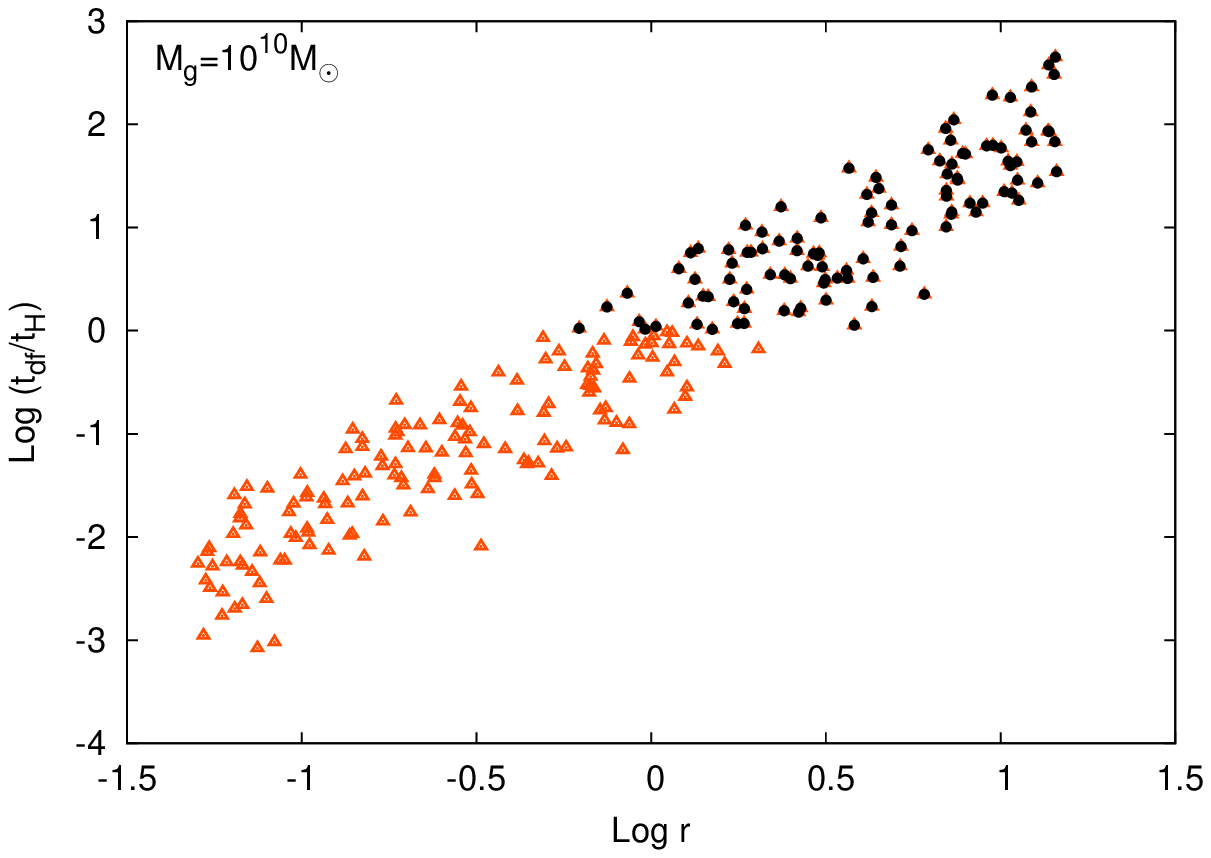}
\includegraphics[width=8cm]{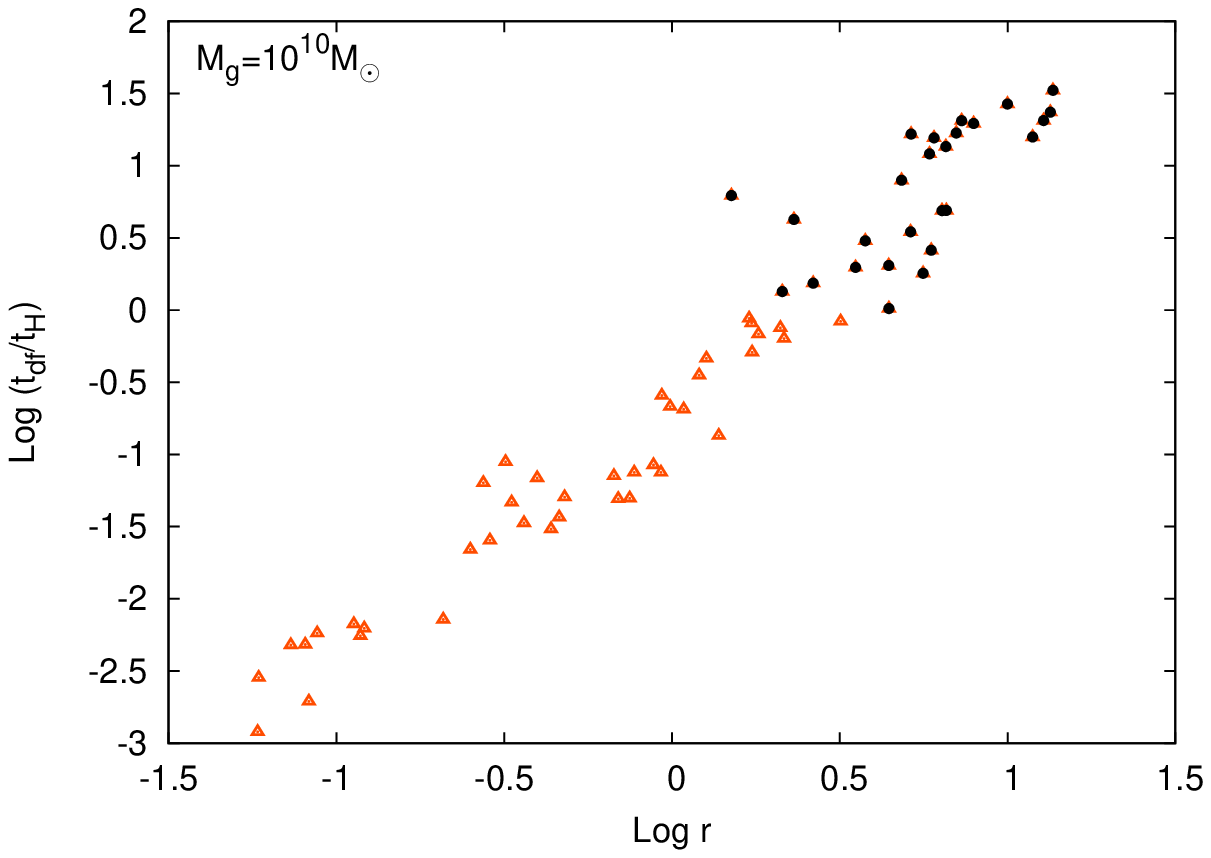}
\includegraphics[width=8cm]{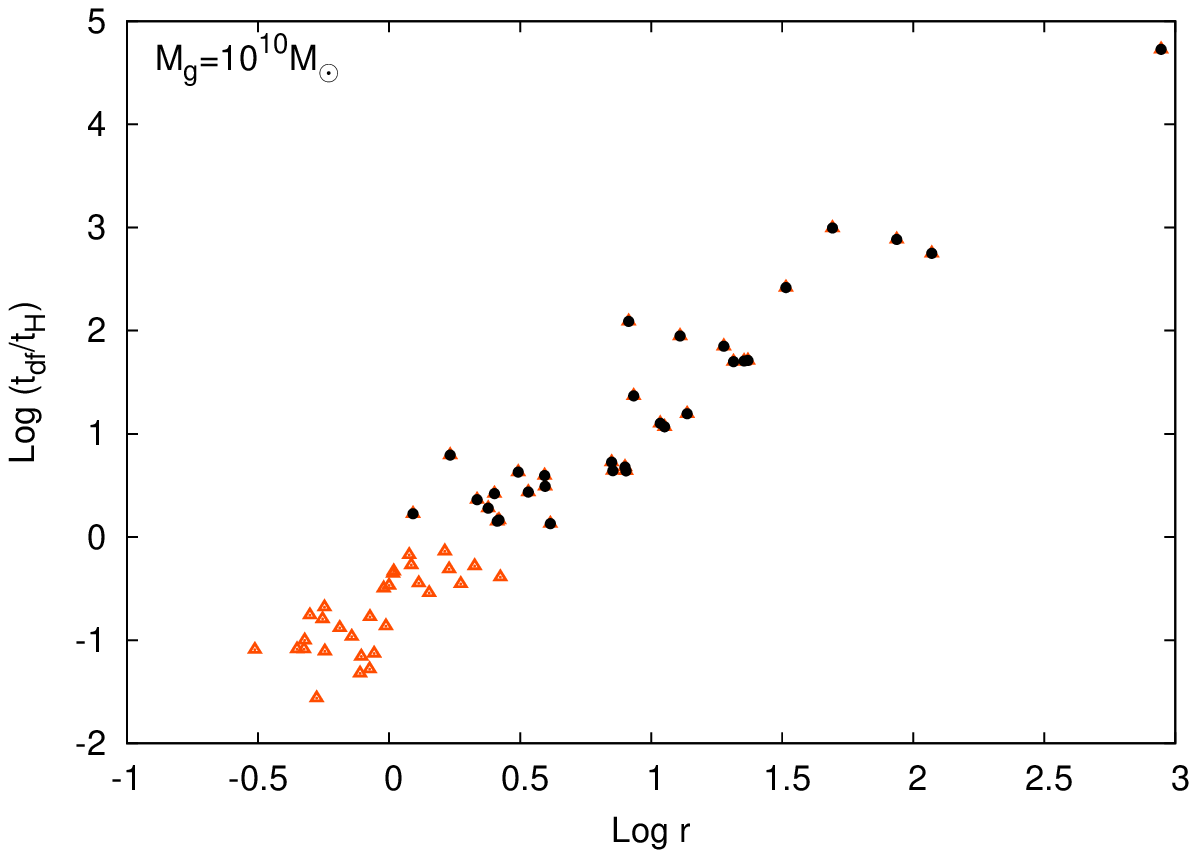}
\includegraphics[width=8cm]{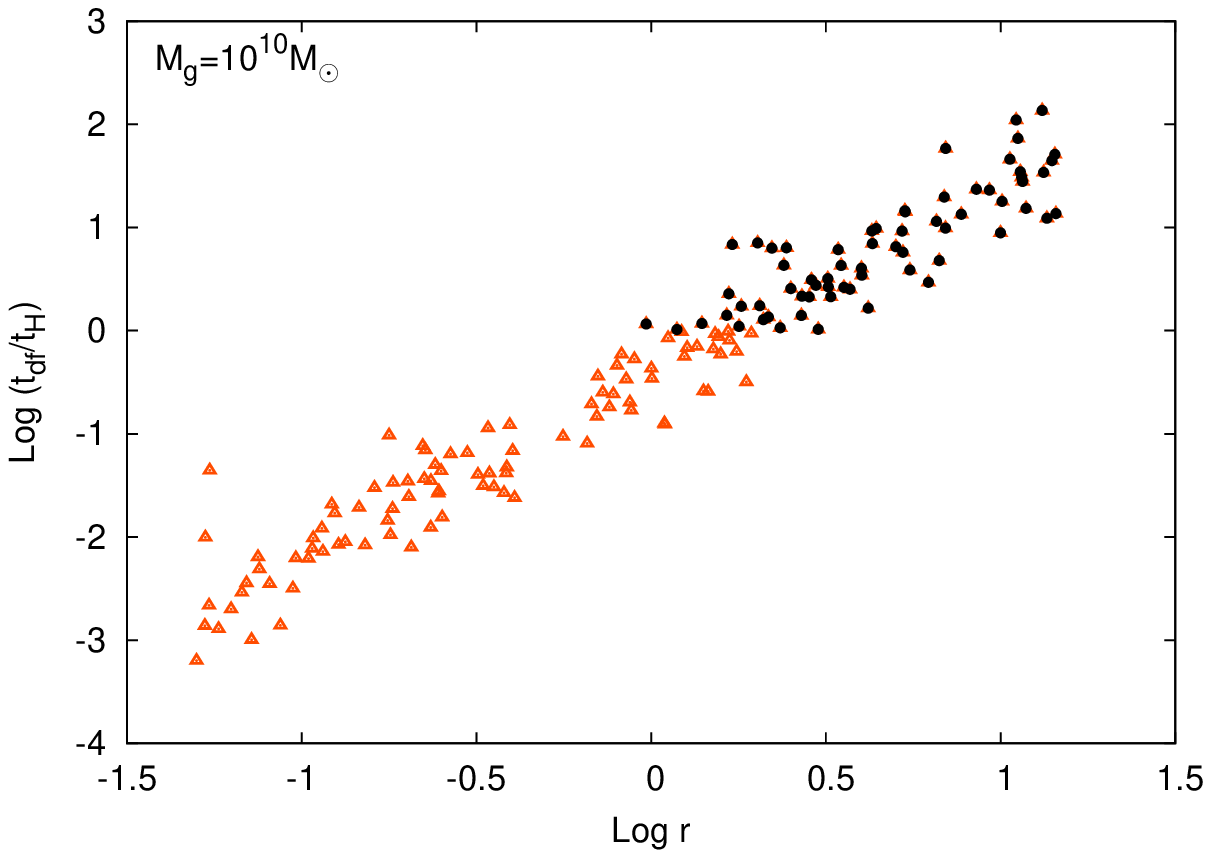}
\caption{From top to bottom, the ratio $t_{\rm df}/t_H$ is shown for a sampled galaxy $M_{\rm g}=10^{10}$M$_\odot$ in PLS, RND, RHO and GSS models, respectively.
Decayed clusters (triangles) lie all in a region whose radius is roughly  $r\sim 3 \-\mathrm{kpc}$.}
\label{tdfDIST}
\end{figure}

\begin{figure}
\includegraphics[width=8cm]{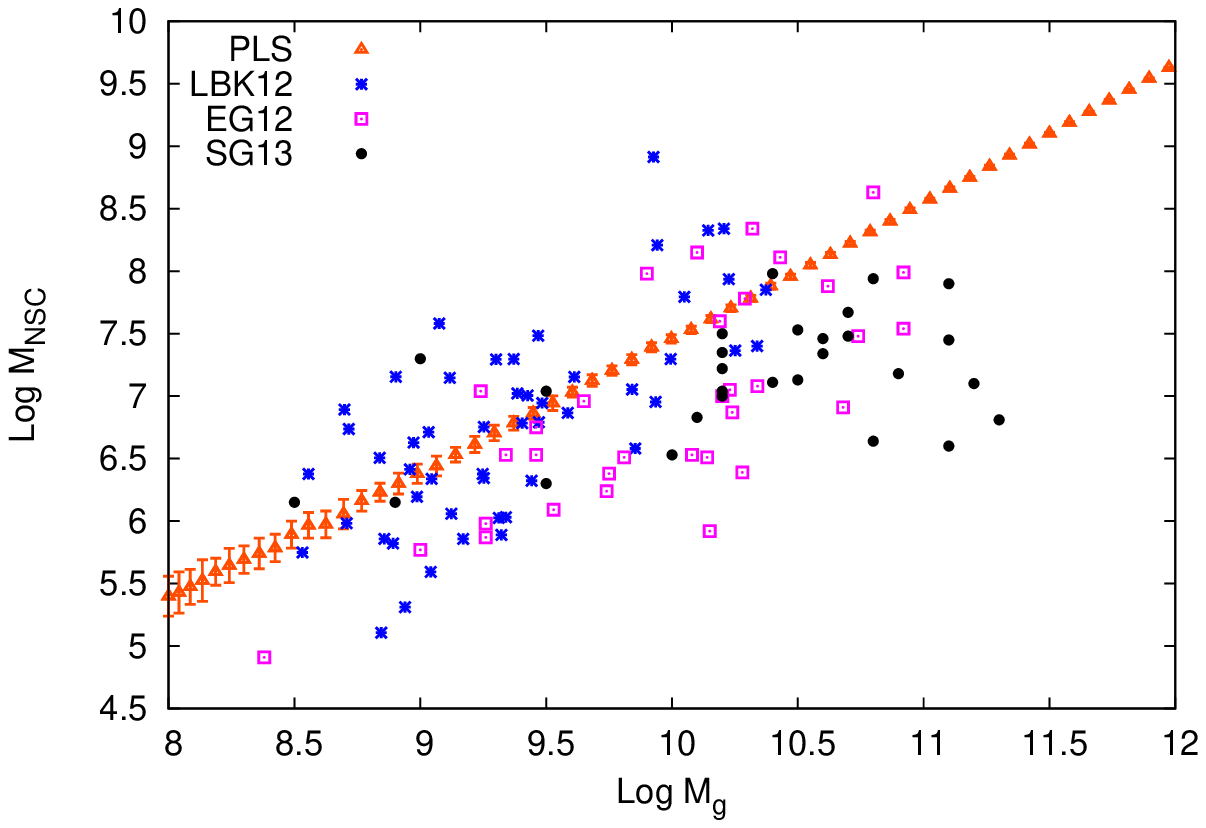}
\includegraphics[width=8cm]{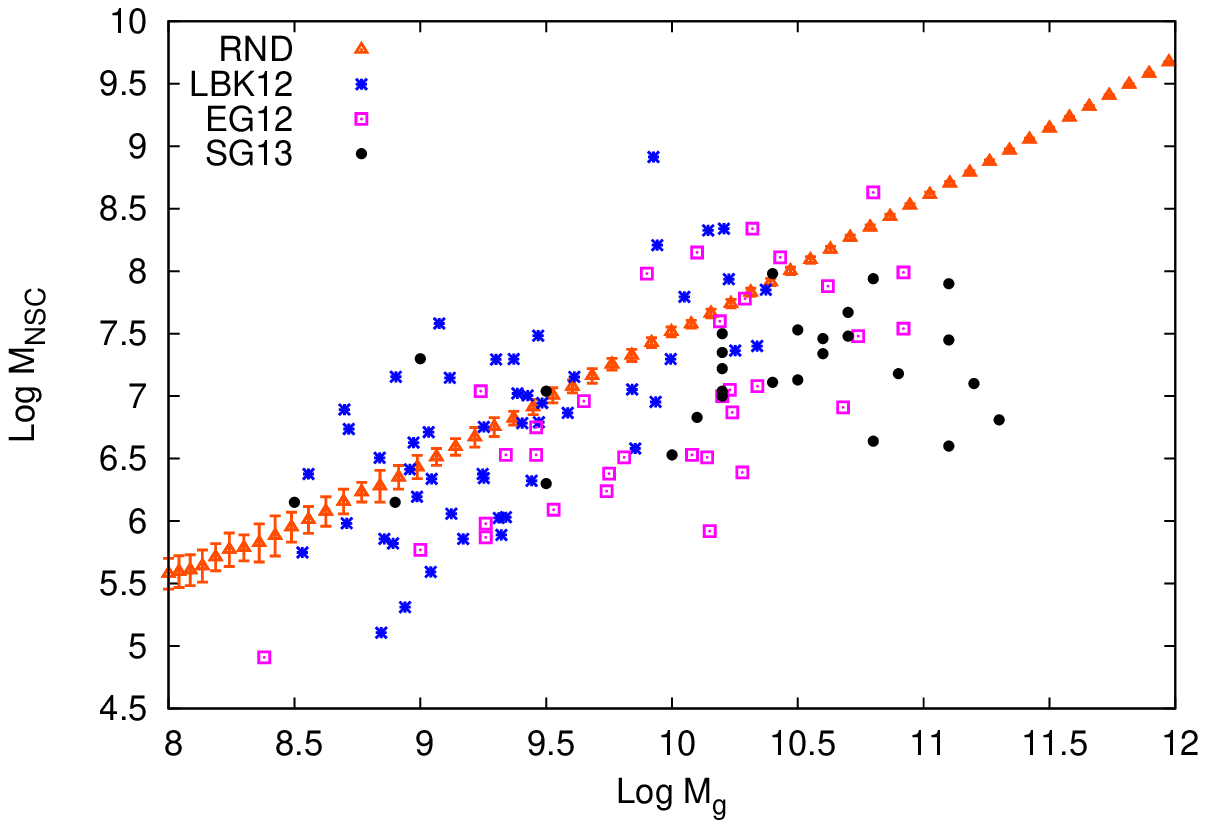}
\includegraphics[width=8cm]{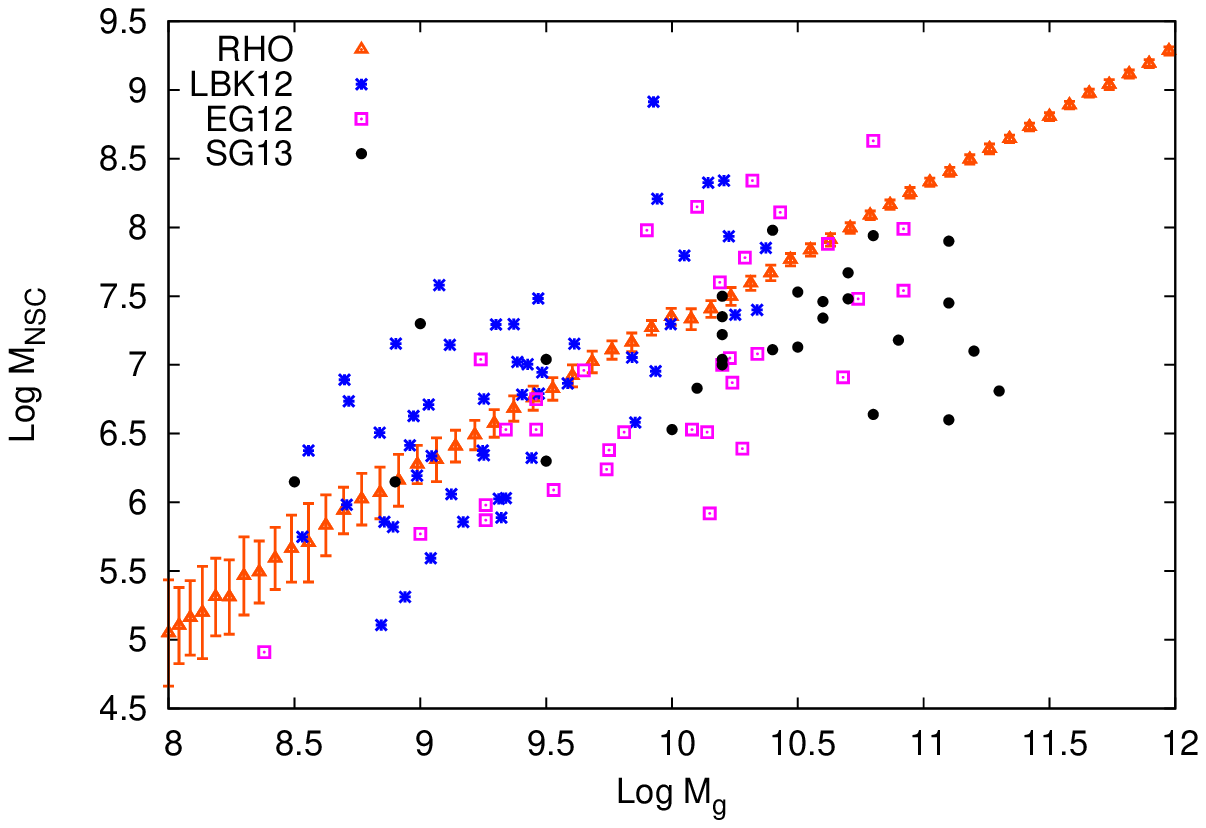}
\includegraphics[width=8cm]{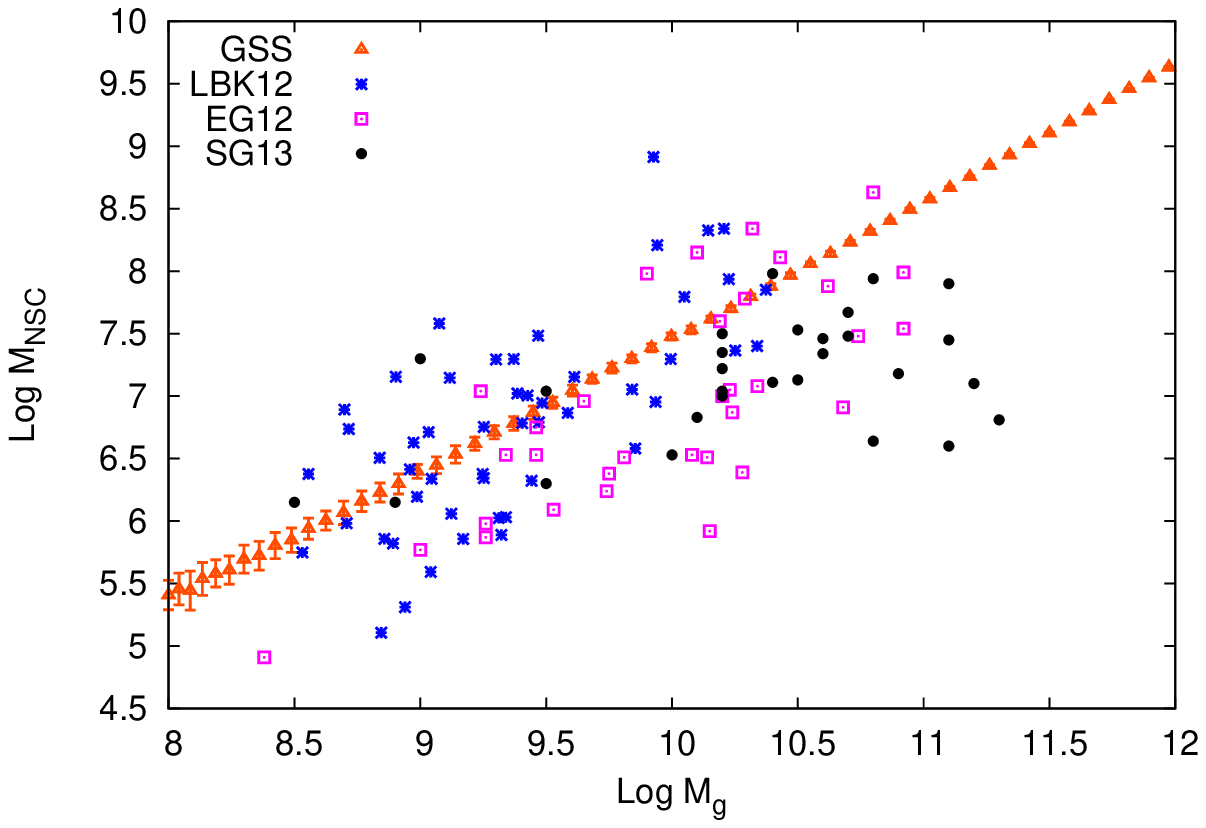}
\caption{NSC masses with respect hosts masses. Predicted values (triangles) are compared with data given in LKB12 (stars), EG12 (squares) and SG13 (filled circles).}
\label{MnscMg}
\end{figure}

Figures \ref{tdfDIST} show the ratio between the df decay time and the Hubble time for GCs 
belonging to a galaxy of mass $M_{\rm g}=10^{10}$M$_\odot$ whose GCS is sampled with the PLS, RND, GSS and RHO models. All clusters with $t_{\rm df}/t_H<1$ are decayed, therefore, the expected NSC mass is evaluated by summing the masses of all the decayed clusters, and in Figure \ref{MnscMg} the resulting NSC mass vs. the host mass is reported.

However, it is not easy to understand which model fits better the observations. A more quantitative analysis is needed to  
reveal the real agreement, as, for instance, that of drawing scaling laws which connect the NSC mass with some of the host properties

In the following Section, we deepen the study of the comparison of our 'theoretical' and observed NSC, drawing the 'theoretical' scaling laws mentioned above to compare them with those actually observed.
and actually observed scaling laws.

\section{Scaling laws}
\label{laws}

As we said in the Introduction, the existence of correlations and scaling relations between the central compact object and the galactic host parameters may be an important clue to the understanding of the actual mechanisms of CMO formation.

It is well known that SMBH masses show a tight correlation with the host galaxy bulge velocity dispersion, $\sigma_{\rm g}$, \citep{frrs} and with the galactic bulge mass, $M_b$, (see for example \cite{marconi} and \cite{harix}). The implication claimed is that similar processes drove both SMBH and galaxy growth. In particular, \cite{silk} suggested that a feedback  exists between the early stage of life of a galaxy and its central BH.

In the last years, many studies were devoted to derive, also, scaling relations among NSCs and their host galaxies, finding that they follow relations in part similar as SMBHs do \citep{rossa}. However, it is still unclear what, if any, the two different types of CMOs have in common, so to imply an intimate link between central galactic BHs and NSCs growth and evolution. Actually, differences in scaling relations of BHs respect to NSCs are being presently debated.
As an example, \cite{frrs} claimed that NSCs follow the same mass-sigma relation of massive central BHs, which is a power law with an exponent between $4$ and $5$. On the other side,  \cite{Graham} and \cite{LGH} find a significantly shallower relation for the mass-$\sigma$ relation of NSCs, with the exponent in the range $1.52$ to $3$. 

At this regard, while the 'in situ" model is compatible with the steeper relation found in \cite{frrs} (see for example \cite*{McLgh}), the 'dry merger" scenario, instead, fits well with the \cite{Graham} and \cite{LGH} relations, as we have seen in Section \ref{teo} of this work (see also \cite{Ant13}).

\begin{figure}
\includegraphics[width=8cm]{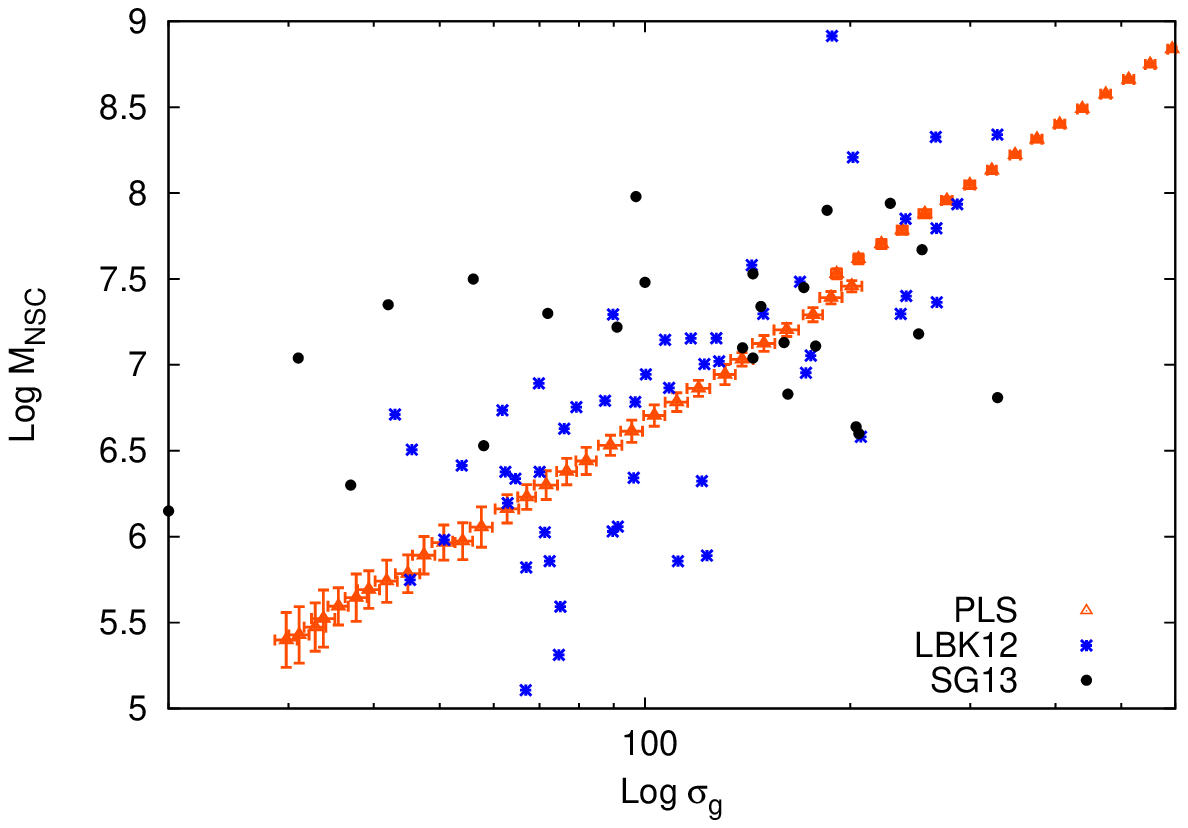}
\includegraphics[width=8cm]{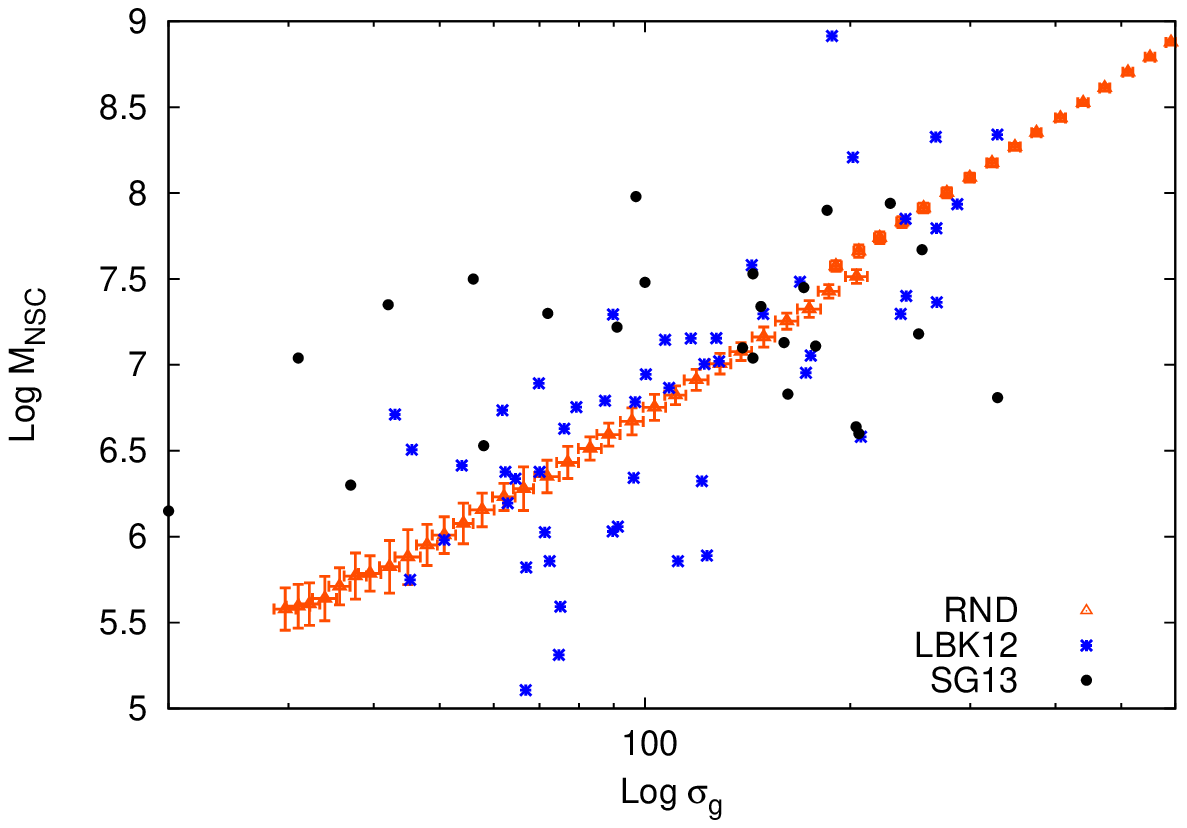}
\includegraphics[width=8cm]{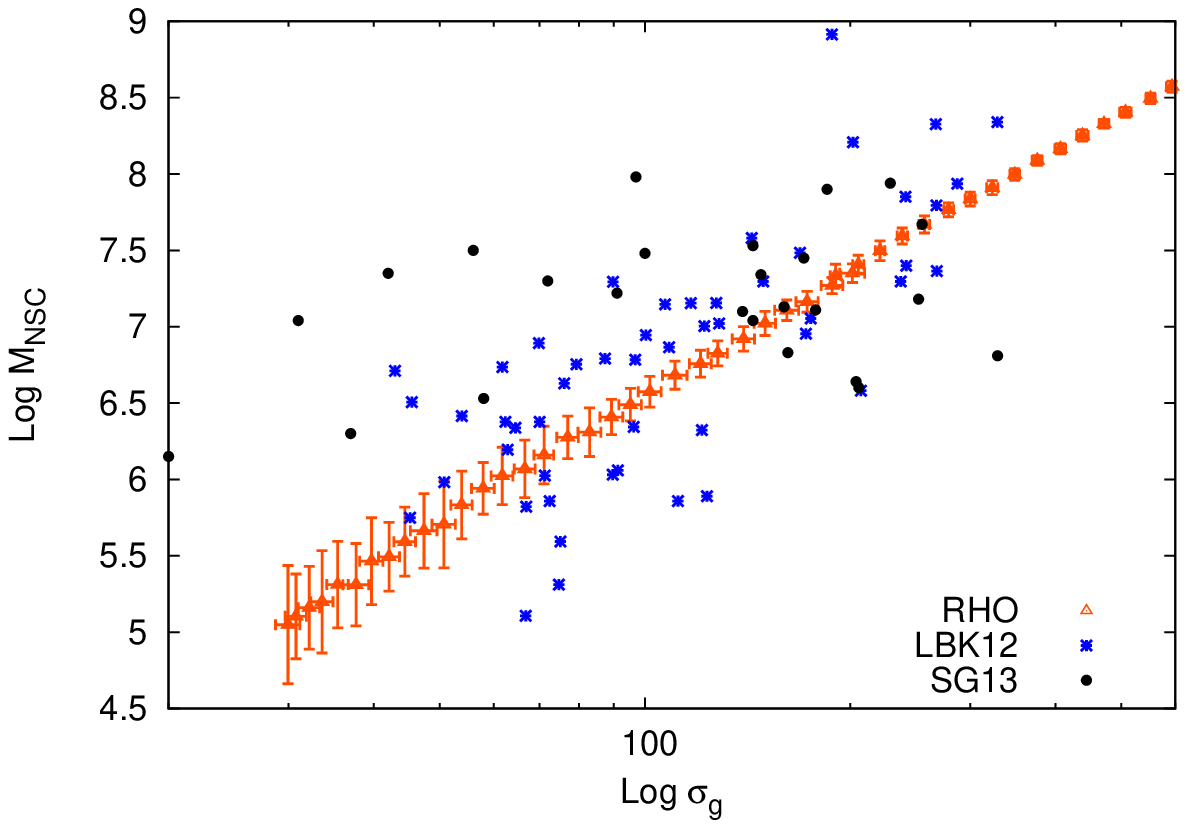}
\includegraphics[width=8cm]{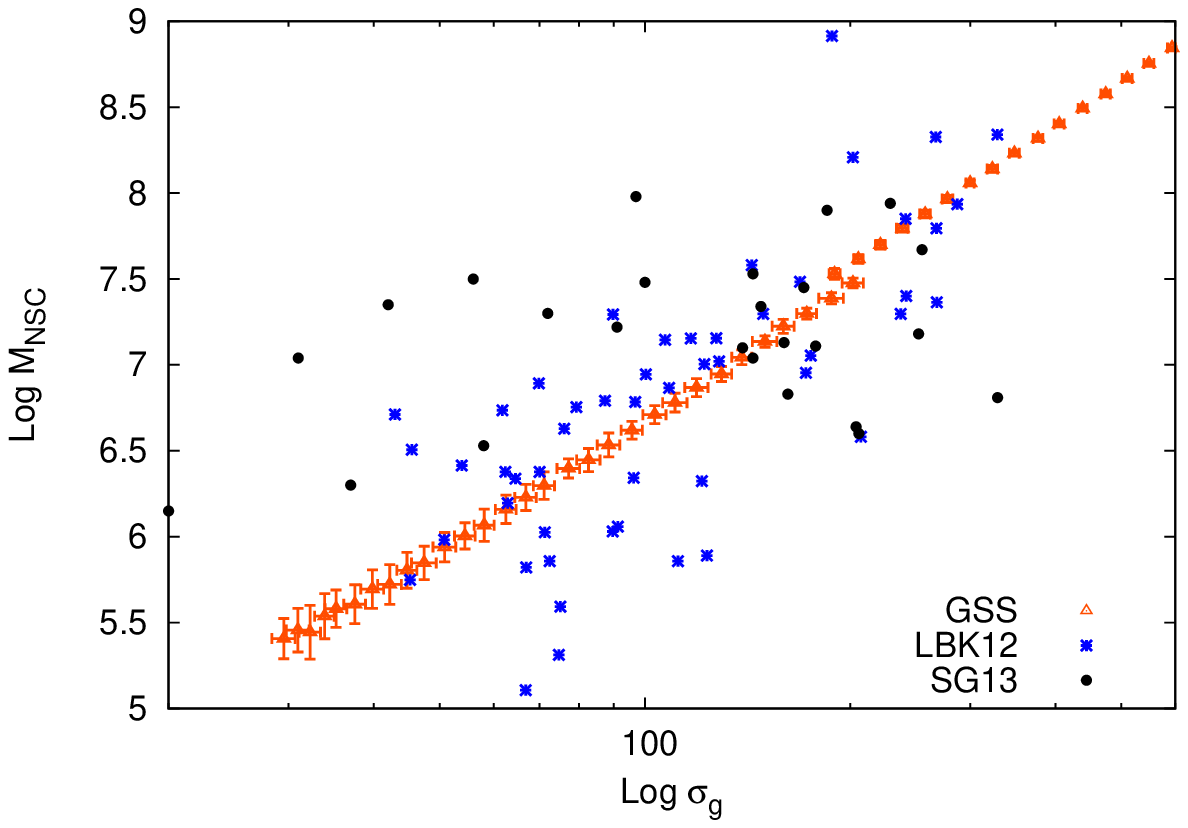}
\caption{NSC masses vs. hosts velocity dispersion. Predicted values (triangles) are compared with data given in LKB12 (stars) and SG13 (filled circles).}
\label{MnscSig}
\end{figure}

\begin{figure}
\includegraphics[width=8cm]{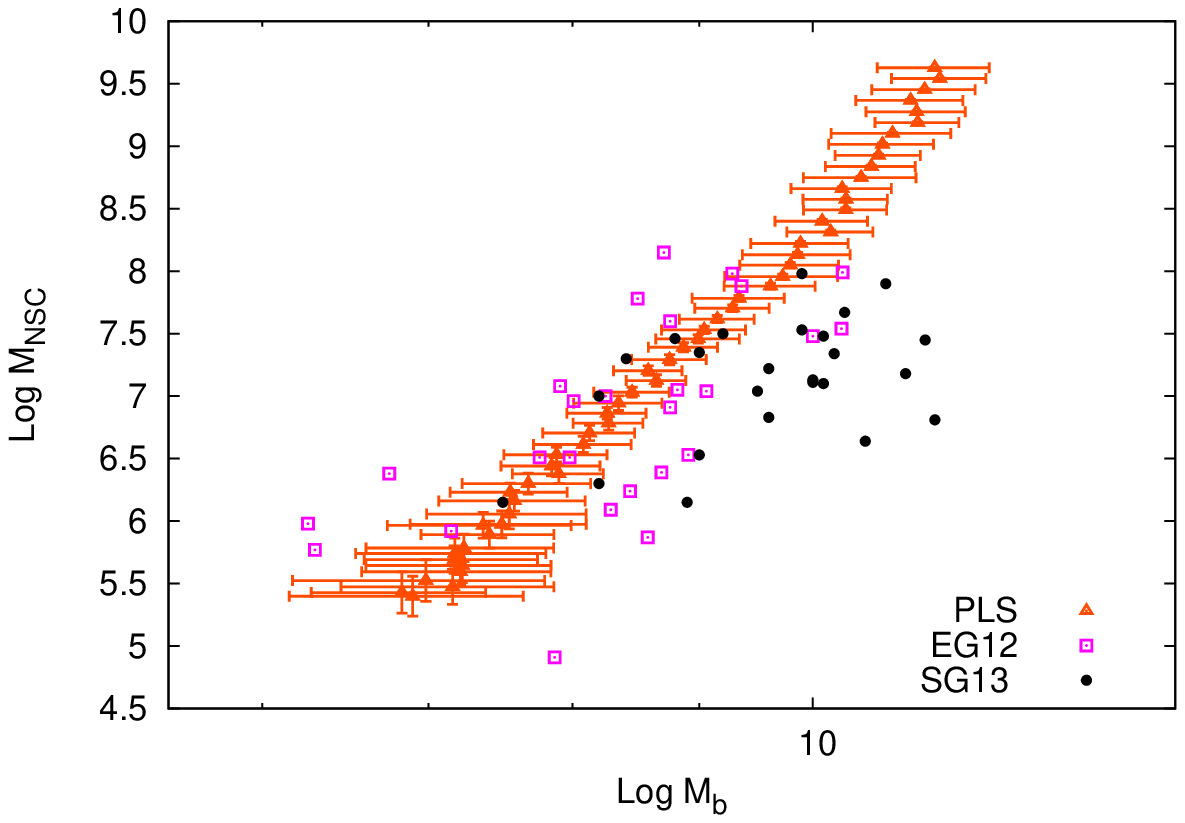}
\includegraphics[width=8cm]{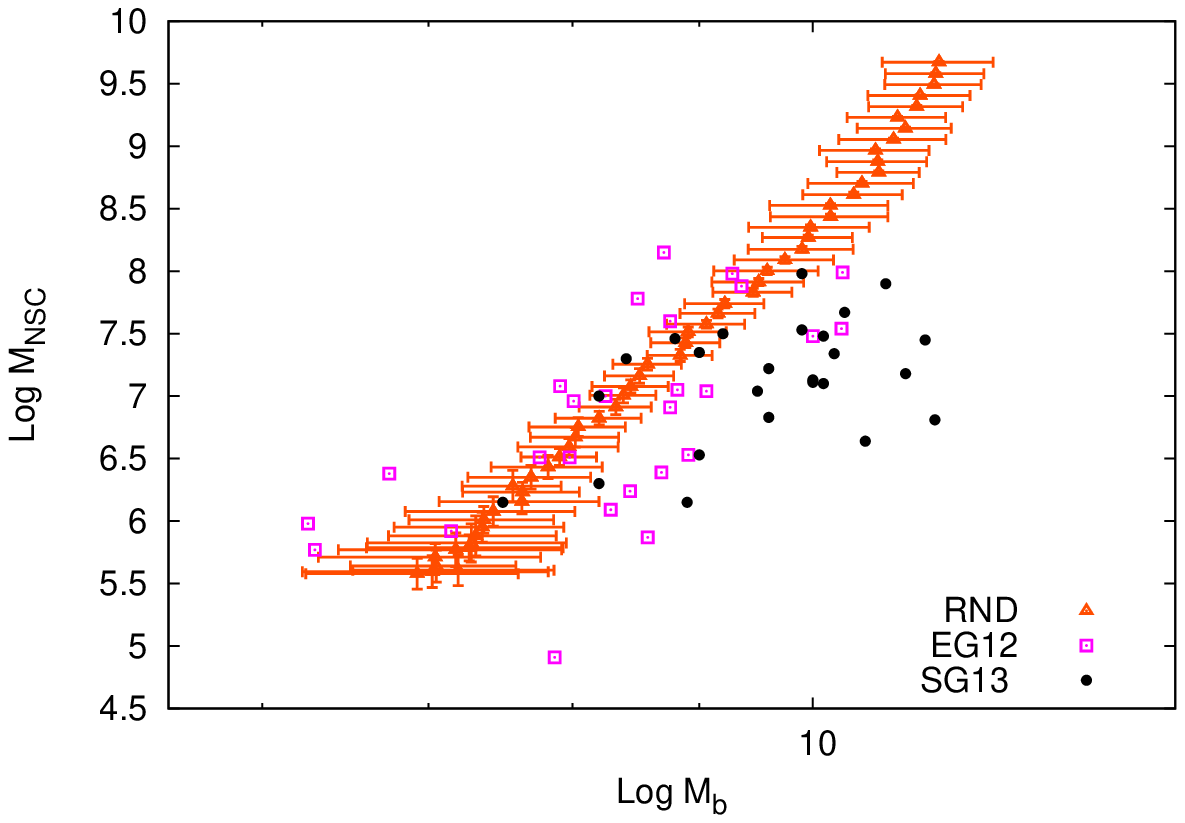}
\includegraphics[width=8cm]{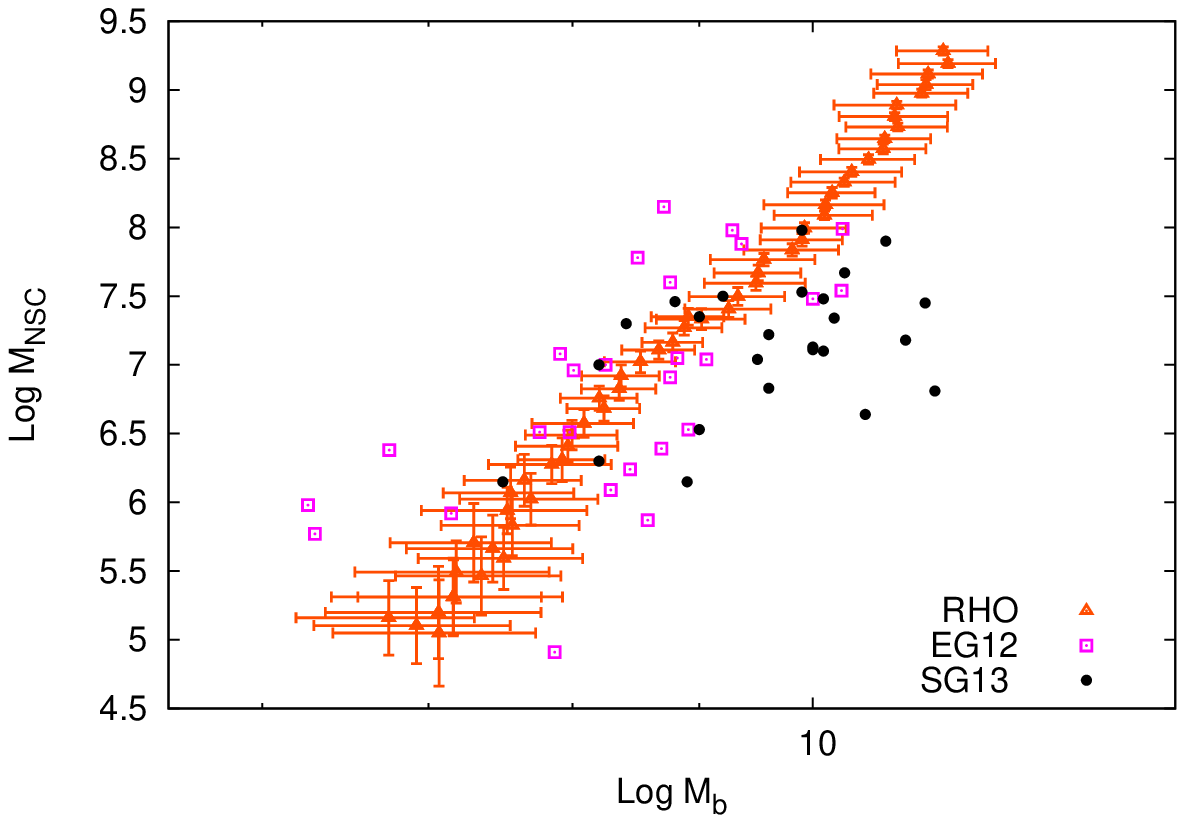}
\includegraphics[width=8cm]{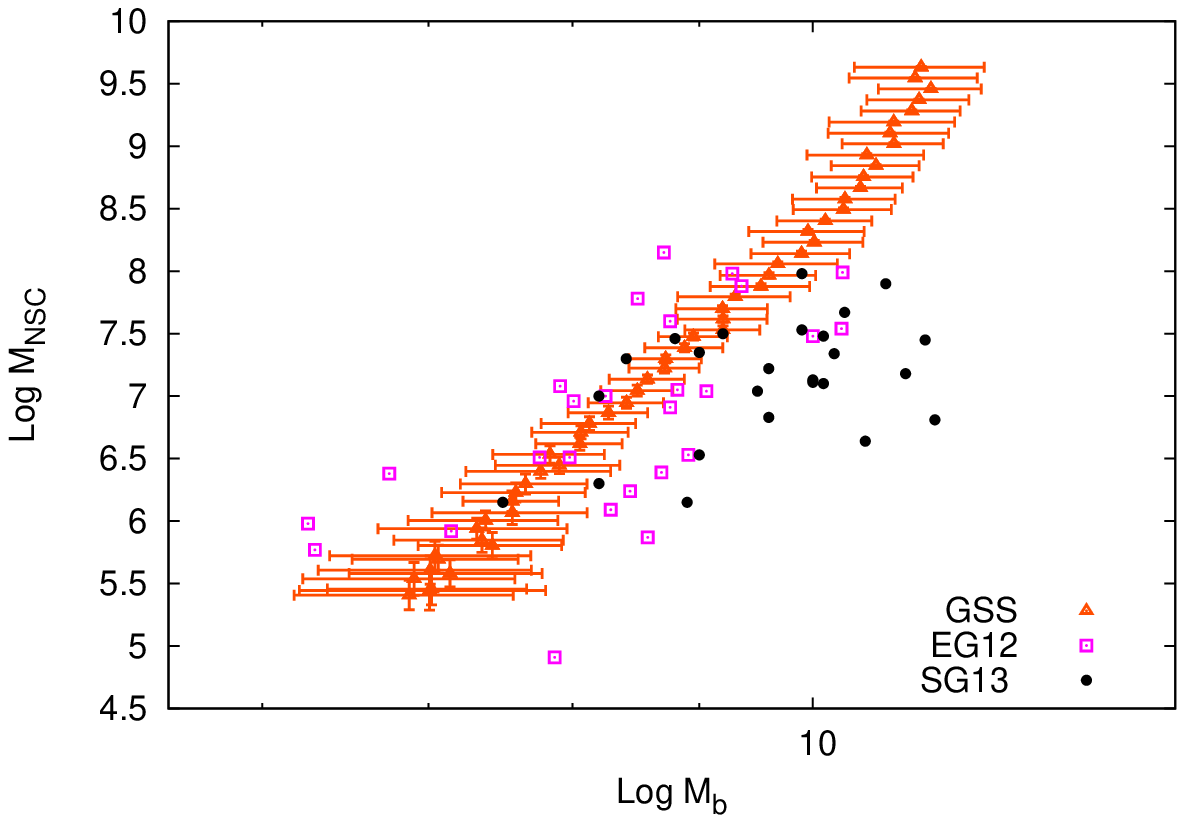}
\caption{NSC masses with respect hosts bulge masses. Predicted values with errors (triangles) are compared with data given EG12 (squares) and SG13 (filled circles).}
\label{MnscMb}
\end{figure}

By means of both the statistical and the analytical approaches presented above we can draw various correlations, including $M_{\rm NSC}-M_{\rm g}$ as well as $M_{\rm NSC}-M_{\rm b}$ and $M_{\rm NSC}-\sigma_{\rm g}$ relations. 
At this scope, in Figure \ref{MnscSig} and Figure \ref{MnscMb} we report, respectively, the NSC masses as functions of the host galaxy velocity dispersion and bulge mass of our models compared with observed data. 
Note that due to the somewhat ill definition of the bulge, the relation $M_{\rm NSC}$ vs $M_{\rm b}$ is  not very reliable.

\subsection{$\mathbf{\mathit{M}_{\rm NSC}-\mathit{M}_{g}}$ relation}

It has been shown that the NSC mass correlates better with the total galaxy mass, while the correlation with the bulge is not statistically very significant \citep{ERWGD}.

We obtained a power law best-fitting for our sampled correlation between $M_{\rm NSC}$ and $M_{\rm g}$ in the form
\begin{equation}
 {\rm Log}\left(\frac{M_{\rm NSC}}{10^{7.6}\mathrm{M}_\odot}\right)=a+b{\rm Log}\left(\frac{M_{\rm g}}{10^{9.7}\mathrm{M}_\odot}\right),
\end{equation}

where the coefficients $a$ and $b$ (see Table \ref{tabM}) are computed by using the Marquardt-Levenberg nonlinear-regression algorithm. For the sake of comparison we report Table \ref{tabMobs} the slope $b$ of the best-fittings to the $M_{\rm NSC}-M_{\rm g}$ relations in the EG12, LKB12 and SG13 data samples. 

The comparison between values in Table \ref{tabM} and \ref{tabMobs} indicates that both the analytical ($s=2$) model and the statistical approaches (PLS, GSS, RND and RHO models) give a slope in good agreement with the observed relation, within the errors.

\begin{table}
\caption{$M_{\rm NSC}-M_{\rm g}$ scaling relation parameters for various models.}
\label{tabM}
\begin{center}
\begin{tabular}{ccccc}
\hline
Model & $b$ & $\epsilon_{r_b}$ & $a$ & $\epsilon_{r_a}$ \\
\hline 
s=2 &$0.955$&$0.015$&$-0.206$&$0.021$\\
PLS &$1.0682$&$0.0030$&$-0.4415$&$0.0037$\\
RND &$1.0482$&$0.0051$&$-0.3778$&$0.0062$\\
GSS &$1.0707$&$0.0028$&$-0.4391$&$0.0034$\\ 
RHO &$1.0488$&$0.0061$&$-0.6525$&$0.0075$\\ 
\hline
\end{tabular}
\end{center}
\medskip
	 Column 1: model name as explained in Section \ref{model}. Column 2-5: slope $b$ and zeropoint $a$ and relative errors.
\end{table}

\begin{table}
\caption{$M_{\rm NSC}-M_{\rm g}$ scaling relation parameters as given in literature.}
\label{tabMobs}
\begin{center}
\begin{tabular}{ccccc}
\hline
Model & $b$ & $\epsilon_{r_b}$\\
\hline 
LKB12&$ 1.18$&$ 0.16$\\
EG12 &$ 0.90$&$ 0.21$\\
SG13 &$ 0.88$&$ 0.19$\\
\hline
\end{tabular}
\end{center}
\medskip
	 Column 1: reference paper name. Column 2-3: slope $b$ and and relative error.
\end{table}

\subsection{$\mathbf{\mathit{M}_{\rm NSC}-\mathit{M}_{b}}$ relation}

Using the estimate of the bulge masses given by Equation \ref{MbMg}, the slopes of the logarithmic correlations between $M_{\rm NSC}$ and $M_b$ for our various models are given in Table \ref{tabB}. They compare with the slope of the observational  law of SG1 data sample, which is $b=0.88\pm 0.19$, in agreement, within the error bar, with all theoretical predictions.

\begin{table}
\caption{$M_{\rm NSC}-M_{B}$ relation parameters for various models.}
\label{tabB}
\begin{center}
\begin{tabular}{ccccc}
\hline
Model & $b$ & $\epsilon_{r_b}$ & $a$ & $\epsilon_{r_a}$ \\
\hline 
s=2 &$0.732$&$0.028$&$1.36$&$0.27$\\
PLS &$0.971$&$0.019$&$-1.37$&$0.20$\\
RND &$0.960$&$0.017$&$-1.22$&$0.19$\\
GSS &$1.000$&$0.028$&$-1.61$&$0.30$\\ 
RHO &$0.840$&$0.012$&$-0.31$&$0.13$\\ 
\hline
\end{tabular}
\end{center}
	\medskip
	 Column 1: model name as explained in Section \ref{model}. Column 2-5: slope $b$ and zero-point $a$ with relative errors.
\end{table}

\subsection{$\mathbf{\mathit{M}_{\rm NSC}-\sigma_{\rm g}}$ relation}

The correlation between the NSC mass and the host galaxy velocity dispersion is probably the most interesting correlation to analyse, because it can give useful hints about relations between the two types of CMOs (SMBHs and NSCs).
If for NSCs and SMBHs a similar mass-sigma correlation holds, one could infer that they shared the same evolutionary path. 

Actually, our theoretical results point towards a weak scaling of the NSC mass with $\sigma_{\rm g}$. 
As shown in Tables \ref{tabS} and \ref{tabSobs}, all our theoretical models give a slope for the mass-sigma relation $2< b_\sigma < 3$, in good agreement with that obtained by observations and just slightly larger than that obtained with the simple dynamical friction based analytical considerations in Sect. \ref{teo}. Given that the SMBHs mass depends more strongly on $\sigma_{\rm g}$, this result would likely imply that NSCs and SMBHs  do not share the same evolutionary history, or, at least, that some different kind of interaction between the two types of objects and the background occurred.
\begin{table}
\caption{$M_{\rm NSC}-\sigma_{\rm g}$ relations parameters for various models.}
\label{tabS}
\begin{center}
\begin{tabular}{ccccc}
\hline
Model & $b$ & $\epsilon_{r_b}$ & $a$ & $\epsilon_{r_a}$ \\
\hline 
s=2 &$2.410$&$0.036$&$-0.336$&$0.021$\\
PLS &$2.699$&$0.015$&$-0.5816$&$0.0072$\\
RND &$2.649$&$0.019$&$-0.5139$&$0.0093$\\
GSS &$2.705$&$0.014$&$-0.5781$&$0.0067$\\ 
RHO &$2.651$&$0.012$&$-0.788$&$0.012$\\ 
\hline
\end{tabular}
\end{center}
	\medskip
	 Column 1: model name as explained in Section \ref{model}. Column 2-5: slope $b$ and zero-point $a$ and relative errors. The relation used is:$\log(M_{\rm NSC}/10^{7.6}\mathrm{M}_\odot)\propto \log(\sigma_{\rm g}/200kms^{-1})$.
\end{table}
\begin{table}
\caption{$M_{\rm NSC}-\sigma_{\rm g}$ relations parameters from the literature.}
\label{tabSobs}
\begin{center}
\begin{tabular}{ccccc}
\hline
Model & $b$ & $\epsilon_{r_b}$\\
\hline 
LKB12&$2.73$&$0.29$\\
SG13 &$2.11$&$0.31$\\
\hline
\end{tabular}
\end{center}
	\medskip
	 Column 1: sample name. Column 2-3: slope $b$ and and relative error.
\end{table}

As final remark of this Section, we note that the Fig. 16 of \cite{rossa} paper shows a relevant feature that NSC formation model should interpret. On one side it gives evidence that much of the mass of NSC is in old stars (thing straightforwardly compatible with the merger model); on the other side, it seems to indicate the presence of older star population in more massive NSCs. This is not at odd with the merger model. Actually, there is evidence of the presence of a certain fraction of young stars in NSCs (see for instance the Milky Way NSC) and this implies that some star formation occurred there in relatively recent imes from some gas there present. Given this, if we assume that the quantity of newly born (in situ) stars is the same in different, increasing in mass, galaxies hosting
more massive NSCs, it comes back naturally that these more massive NSCs are, indeed, more massive because grown by a larger quantity of mass in decayed globular clusters which, consequently, have an increasingly old stellar population inside, due to the increased number fraction of old to young stars. Anyway, this is only a speculation that deserves a deeper investigation.

\section{Tidal disruption effects}
\label{tid}

In the previous Sections we showed that the dry-merger scenario provides scaling relations connecting the NSC masses with global parameters of their hosts. However, there are at least two effects which could prevent the formation of NSCs, 
acting in competition with the dynamical friction process: the two-body relaxation mechanism and the tidal heating process.
In this Section we study their effects on the formation of NSCs and show that the scaling laws derived in this case still agree with observations in the whole range of galaxy masses. 

In the last section, we neglected the effect of the disruption of cluster since, as we will show in this section, in small galaxies ($M_{\rm g}\leqslant 10^{10}$M$_\odot$) the dominant process in the formation of the galactic nucleus is the dynamical friction process, while in heavier galaxies tidal processes could prevent its formation.
 
During its lifetime, a GC undergoes internal dynamical evolution experiencing two-body relaxation and suffering of external tidal perturbations that, in some cases, can lead to its total, or partial, dissolution. 
Actually, it is well known that two-body encounters between stars may bring some of them beyond the GC tidal boundary after few hundred times the typical two-body relaxation time \citep{spit87}. 
\cite*{GLB08}, using results by \cite{bmgrt01}, gave the following formula for the evaluation of the dissolution time of a cluster due to the effects of two-body encounters: 

\begin{equation}
\tau_{dis} (\mathrm{Gyr})=\left(\frac{M}{10^4\rm{M}_\odot}\right)^{0.62}\left(\frac{r}{\mathrm{kpc}}\right)\left(\frac{v_{\rm c}}{220\mathrm{kms^{-1}}}\right)^{-1}(1-e),
\label{Tdrel}
\end{equation} 

where $M$ is the GC mass, $r$ is the distance from the galactic centre, $v_{\rm c}$ the circular velocity at $r$, and $e$ the eccentricity of the orbit.

Moreover, gravitational encounters between the stellar system and a perturber (which could be a black hole), the disc,
or the nucleus, of the galaxy, could lead to the destruction of the system over a time comparable to the dynamical friction decay time. This implies that the phenomenon of cluster destruction cannot be neglected.
In the case in which the perturber is a point mass, a black hole, \cite{spitzer58, spit87} studied the effect of such perturbation on a stellar system of mass $M_{\rm s}$, in the hypothesis that the duration of the encounter is short compared to the internal crossing time of the cluster. This is the impulse approximation, which assumes that as a consequence of the encounter, stars in the perturbed system suffer only a change in their velocities, but not in the positions. Moreover, due to the slow duration of the perturbation, the cluster trajectory can be approximated with a straight line.
Under this hypothesis, it is possible to show that the cluster, as a consequence of the gravitational encounter with the perturber of mass $M_{\rm p}$, gains an energy per unit mass:
\begin{equation}
\Delta E= \frac{4G^2M_{\rm p}^2}{3V^2b^4}\left\langle R^2 \right\rangle,
\end{equation}

where $V$ is the relative velocity between the two objects, $b$ is the impact parameter and $\sqrt{\left\langle R^2 \right\rangle}$ is the mean dimension of the perturbed cluster.

A number of studies have been devoted to generalize this result to an extended spherical perturber with an
arbitrary mass distribution \citep*{AW,GHO,GO}, and to the case where the perturber is a spherical nucleus of stars embedded in a triaxial ellipsoid \citep*{OBS,Dolc93}.

Defining $U(b/r_{\rm h})$ the ratio between the impulsive energy change due to a perturber of half mass radius $r_{\rm h}$ and that caused by a point of same mass, $M_{\rm p}$, the total change in energy per unit mass caused by a mass distribution is given by:

\begin{equation}
\Delta E= \frac{4G^2M_{\rm p}^2}{3V^2b^4}\left\langle r^2 \right\rangle U(b/r_{\rm h});
\label{Edot}
\end{equation}

where the function $U(b/r_{\rm h})$ drops rapidly to $0$ when $b/r_{\rm h}$ approaches zero, while tends to $1$ for large values of $b/r_{\rm h}$ and should be evaluated numerically.

If the energy change exceeds the internal gravitational energy of the system per unit mass \citep{spitzer58}:
\begin{equation}
E=\frac{3}{5}\frac{GM_{\rm s}}{R_{\rm s}},
\label{Egra}
\end{equation}
the cluster is disrupted. 
The typical time over which this disruption occurs is
\begin{equation}
t_{\rm dis}=\frac{E}{\Delta E} nT.
\label{tdis}
\end{equation}
where $T$ is the orbital period of the cluster and $n$ is the number of encounters within a period.

Further, the encounters are charcterized by two extreme regimes: the catastrophic regime, if a single encounter could disrupt completely the system, and the diffusive regime, when the cumulative effect of encounters leads to the disruption of the system over a longer time.
Defining as $b_{\rm M}$ the impact parameter that corresponds to an energy enhancement equal to the internal gravitational energy of the
system, i.e. $\Delta E=E$, it is possible to determine the duration of the encounter, $t_{\rm enc}\sim b_{\rm M}/\sigma_{\rm rel}$, that is the typical time-scale which discriminates between the two regimes: hence a catastrophic collision occurs if the duration of the encounter is short compared to the crossing time of the cluster, $t_{\rm cr}\sim R_{\rm s}/V$; on the other hand, slower encounter leads to the diffusive regime.

Therefore, in the case of the catastrophic regime ($t_{\rm enc}<t_{\rm cr}$), the tipical disruption time is given by:
\begin{equation}
t_{\rm cat} \simeq \frac{k_{\rm cat}}{G\rho_{\rm p}}\left(\frac{GM_{\rm s}}{R_{\rm s}^3}\right)^{1/2},
\end{equation}
where $\rho_{\rm p}$ is the perturber density, and $k_{\rm cat}$ a constant.
In the diffusive regime $t_{\rm enc}\geq t_{\rm cr}$, instead, it is possible to show that the disruption time is given by:
\begin{equation}
t_{\rm dif}=\frac{0.043}{W}\frac{\sigma_{\rm rel}M_{\rm s}r_{\rm h}^2}{GM_{\rm p}\rho_{\rm p}R^3},
\end{equation}
where $W$ is defined as:
\begin{equation}
W=\int\frac{U(x)}{x^3}\mathrm{d}x, 
\end{equation}
where $x=b/r_{\rm h}$ and $U(x)$ is defined above, and should be computed numerically.

\begin{figure}
\centering
\includegraphics[width=8.5cm]{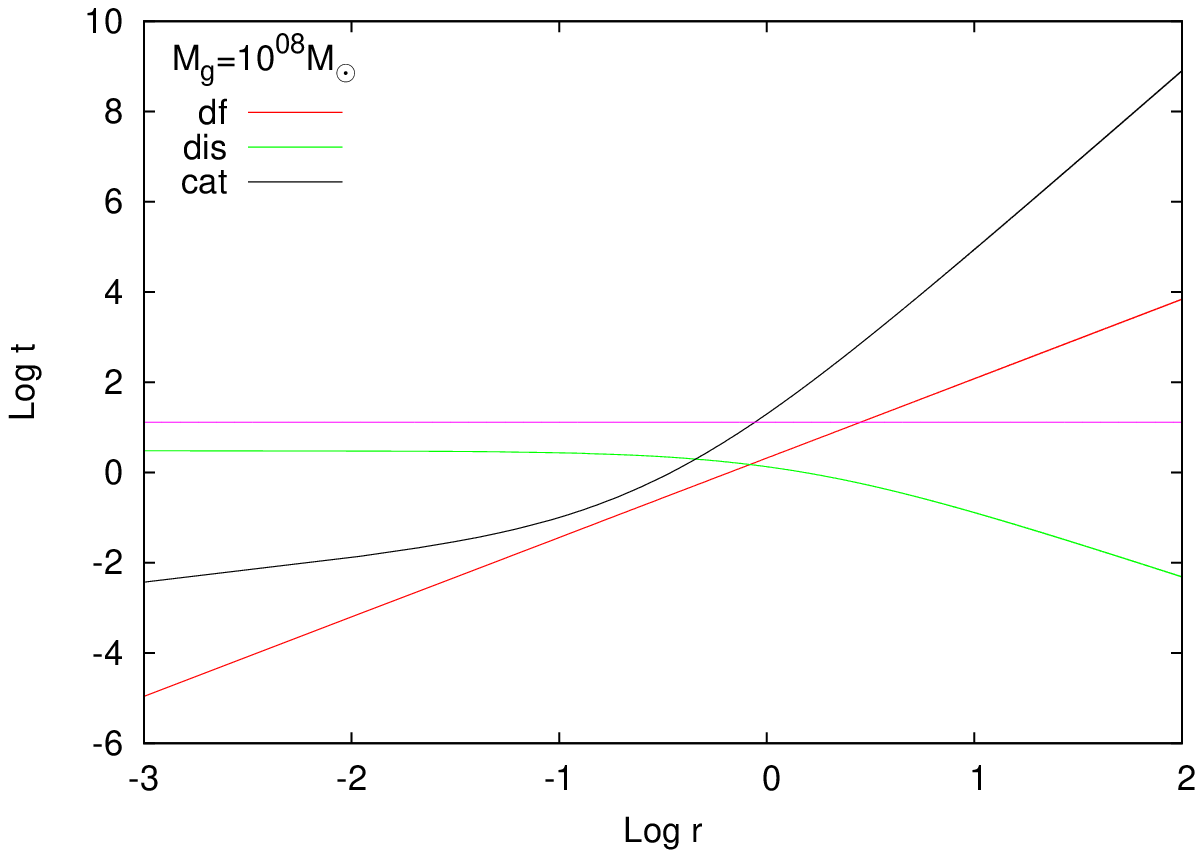} \\
\includegraphics[width=8.5cm]{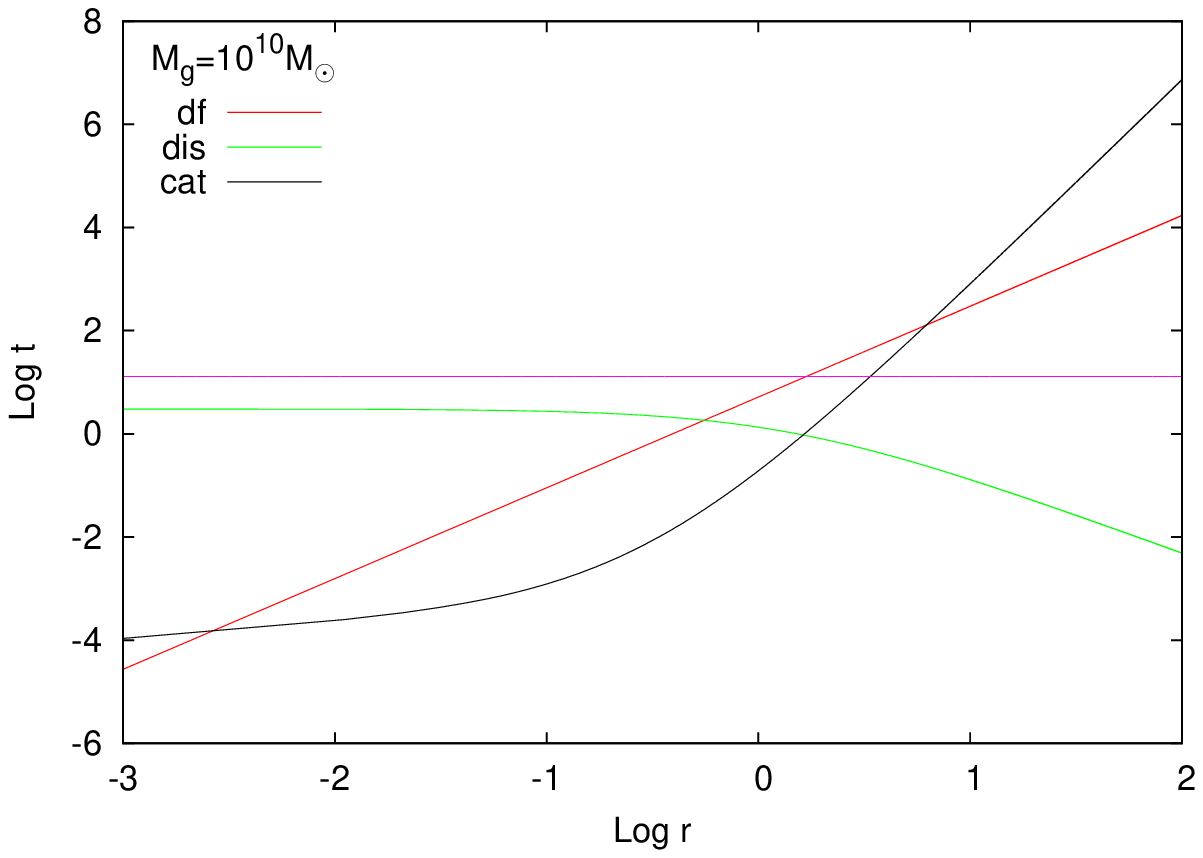} \\
\includegraphics[width=8.5cm]{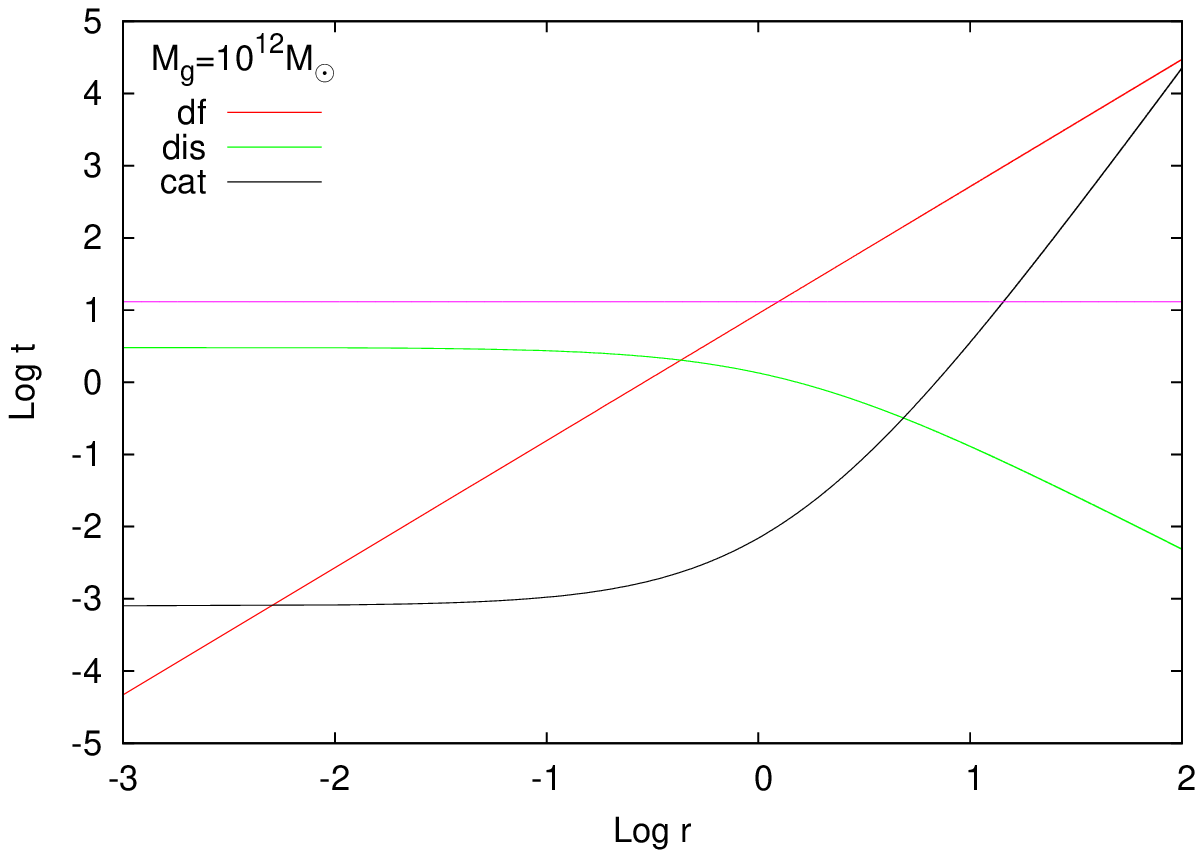}
\caption{Dynamical decay and disruption times for a globular cluster on circular orbit for three different galaxy masses (from top to bottom: $10^8-10^{10}-10^{12}$M$_\odot$). Times are given in units of $1 \-\mathrm{Gyr}$ while positions are given in units of $1 \-\mathrm{kpc}$.
In smaller galaxies, friction process dominates on all length scales, while in heavier systems the tidal disruption is much more rapid in a region $1\- \mathrm{kpc}$ around, suppressing the friction mechanism.}
\label{times}
\end{figure}

Tidal effects are accounted for in our calculations in the above framework, so to investigate their role on the expected value of the NSC mass.
In Figure \ref{times} we compare the dynamical friction time, $t_{\rm df}$, evaluated using Equation \ref{tdf}, the dissolution time, $t_{\rm dis}$, given in Equation \ref{tdis}, and the tidal disruption time in the catastrophic regime $t_{\rm cat}$ for three values of the galaxy mass ($M_{\rm g}=10^8,10^{10},10^{12}$M$_\odot$) as a function of the distance, $r$, from the centre of the host galaxy. On the other hand, the disruption time in the diffusive regime, $t_{\rm dif}$, is not reported in the graph since it is sistematically greater than the other time-scales.
We performed the estimation setting the GC mass to $M=10^6$M$_\odot$ and selecting a circular orbit. 
Looking at Figure \ref{times}, it is clear that while in small galaxies ($M<10^{10}$M$_\odot$), the dynamical friction time is smaller than the disruption times over all length scales, in more massive galaxies it dominates only in a region around the centre of the galaxy, while in the range $0.1-100\- \mathrm{kpc}$ dominates the tidal effect due to the interaction between the cluster and the galactic nucleus, suppressing the role of dynamical friction process and, then, the consequent formation of a NSC.

The two competitive processes, make that the number of decayed clusters depends strongly on their space distribution.

\begin{figure}
\centering
\includegraphics[width=8.5cm]{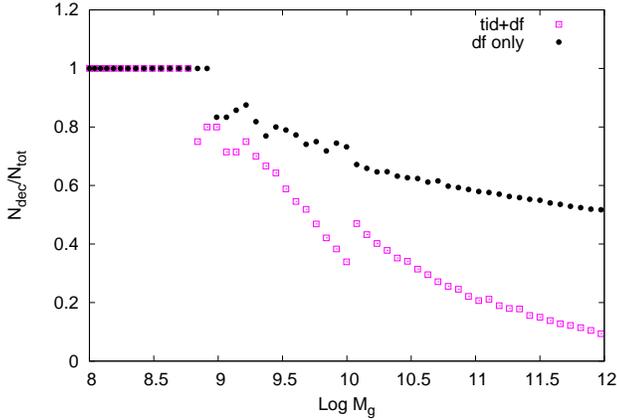}
\caption{The fraction of the number of decayed clusters to their total number considering only df (filled circles) or taking into account the tidal shock mechanism (empty squares) in the RND model. As you can see, the number of decayed clusters in massive galaxies decrease by a factor $5$ when tidal disruption is considered, changing by roughly the same factor the NSC final mass.}
\label{rndN}
\end{figure}

In the RND, PLS and GSS models, we set $d=50$pc as minimum distance from the galaxy centre of the GC sample. Clusters lying in the central $50-100$pc, which are massive enough ($M\gtrsim 10^6$M$_\odot$) or on eccentric orbits are likely to decay and give a large contribute to the final NSC mass. Figure \ref{rndN} shows the fraction of the number of decayed clusters to the total number for the RND model, considering or not the tidal effect.

While in small galaxies almost all the clusters have time to decay, even if the tidal disruption mechanism is active, in massive galaxies the perturbations induced by the external field corresponds to a reduction of the NSC mass.

\begin{figure}
\centering
\includegraphics[width=8.5cm]{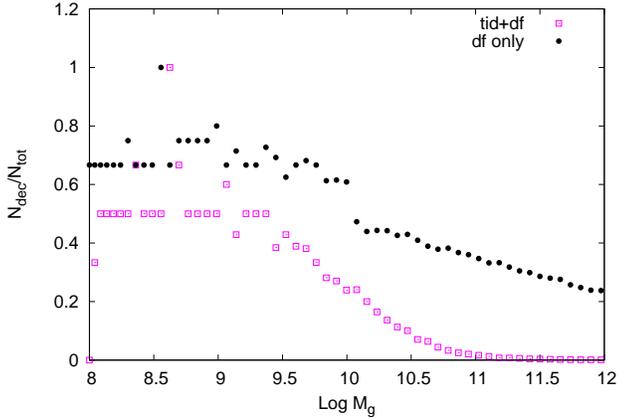}
\caption{Same as in Figure \ref{rndN}, but for the RHO model. In this case, the number of decayed clusters for massive galaxies drops to 0 and there is no NSC formation.}
\label{rhoN}
\end{figure}

However, the action of the tidal disruption mechanism does not change dramatically the final mass of the NSC; in fact, the decrease in mass of the final nuclear cluster is reduced of only the $20\%$ respect to the case when the tidal action
of the galaxy on the clusters is not considered. Hence, the effect of tidal processes in such models is not too important in the determination of the final NSC masses.
On the other hand, since in the RHO model the minimum distance from the galaxy centre allowed for GC formation is given by the constraint that the GCS mass profile follows a Dehnen profile, it could exceed $\sim 50$pc. This implies that many clusters lie in the region 
in which the tidal effects dominate, affecting strongly the final mass of the NSC. In this case, as we can see in Figure \ref{rhoN}, the number of decayed clusters in very massive galaxies drops to zero avoiding the formation of NSCs.

\begin{figure}
\includegraphics[width=7.9cm]{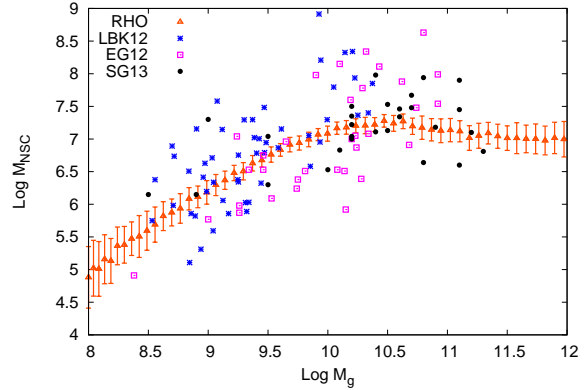}
\includegraphics[width=7.9cm]{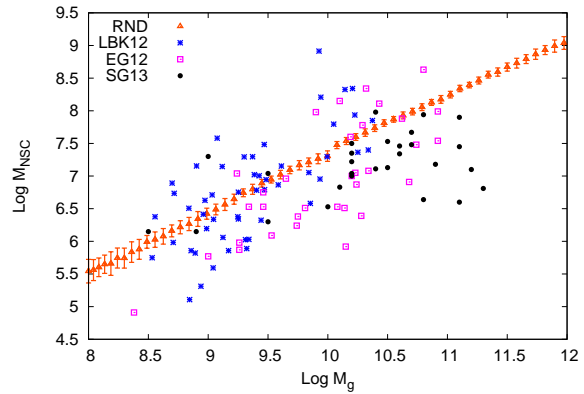}
\includegraphics[width=7.9cm]{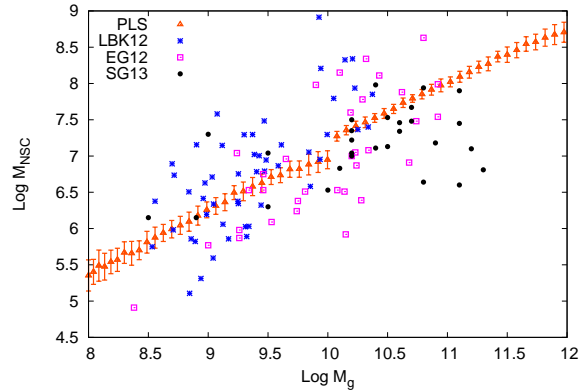}
\includegraphics[width=7.9cm]{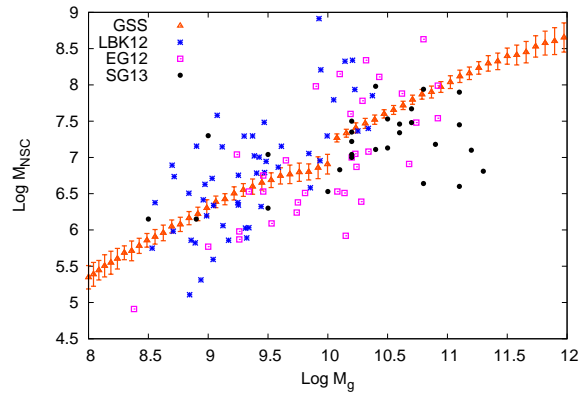}
\caption{NSC masses as a function of the galaxy mass for all the models considered taking in account the tidal disruption processes. In each panel, triangles refer to the results of each model (as indicated in the legend) while stars, squares and filled circles refer to observations.}
\label{RHOtid}
\end{figure}

The difference between the RHO model and the others, where the GCs positions are sampled randomly, puts in evidence two interesting things: the mass distibution of the clusters is not very important in deriving the NSC mass, but instead, what care is how clusters are distributed within the galaxy. This because at intermediate radial scales the disruption time is smaller than the decay time, while it is longer in the central region. Hence, a concentrated spatial distribution allows the formation of the NSC because clusters in the innermost region of the galaxy decay rapidly. 

In Figure \ref{RHOtid} we report the values of NSC masses with respect to host masses for all the models considered taking into account the tidal disruption process. It is seen that in the case of the RHO model the tidal interaction (tidal heating) inibhits the NSC formation for galaxies masses above $3 - 4\times 10^{10}$M$_\odot$, putting in evidence how important the spatial distribution of the clusters is.
It is interesting noting that in this case, the scaling relations found are again in good agreement with the observations, if we restrict the comparison to the actually observed range of masses ($10^8 - 10^{11}$M$_\odot$).
Moreover, it is relevant noting that the flattening observed in Figure \ref{RHOtid} for the RHO model has an observative ``counterpart'': in fact it seems that galaxies more massive than $10^{11}$M$_\odot$ do not host NSCs. Our results suggest that in heavy galaxies, in which tidal processes act against the dynamical friction, should form a NSC lighter than expected. As example, extrapolating from observation the mass of a NSC in a galaxy of $10^{12}$M$_\odot$, we should expect a NSC with $M_{\rm NSC}\sim 10^9$M$_\odot$; however, the range of galaxy masses in which NSCs are observed is dominated mainly by dynamical friction process. Considering instead that in more massive galaxies disrupting processes dominate, it is possible to have NSCs in galaxies heavier than $10^{11}$M$_\odot$ with masses few times $10^7$M$_\odot$. In this picture, NSCs may form in heavy (but not {\it too} heavy) galaxies, but are too small to emerge from the galactic background. In a forthcoming paper, we will investigate such matter in a more complete way.

\section{Summary and Conclusions}
\label{end}

We found that the dry-merger scenario predicts masses for NSCs and scaling laws between them and their host galaxies in excellent agreement with observations. 

The summary of our work is:
\begin{enumerate}
\item an analytical treatment to estimate the formation and growth of NSCs masses has been developed;
\item reliable galaxy models have been provided, as it has been shown comparing theoretical and observational global properties;
\item assuming for the GC system in a galaxy a power-law mass function and a uniform spatial distribution, the analytical predictions fit very well observations (Sect. \ref{res});
\item the consequences of different initial mass distributions of the set of GCs in the host galaxies on the NSC final mass have been investigated from a statistical point of view, by sampling, for each galaxy, its GCS and considering how many clusters were able to sink to the galactic centre within a Hubble time;
\item by means of the statistical approach, we obtained some useful parameters, such as the GC mean mass and the number of survived clusters, which result in good agreement with observations (Sect.\ref{stat} and \ref{statmod});
\item scaling laws which connect the NSC parameters with total mass, velocity dispersion and bulge mass of the host have been deduced; the agreement found between all the models considered and observations indicates that the GC mass distribution does not play a crucial role in determining the final NSC mass (Sect. \ref{laws});
\item the role of tidal disruption mechanism has been investigated under different assumptions for the spatial and mass distributions of GCs in their host galaxies: in RND, PLS and GSS models the tidal heating causes a decrease of predicted NSC masses from few percent in small galaxies (down to $10^{10}$M$_\odot$) to $20\%$ in heavier galaxies. On the other hand, tidal disruption strongly affects NSC formation in the RHO model, where the predicted NSC masses are almost constant in the range $10^{10} - 10^{11}$M$_\odot$.  
An important result is that the best comparison for the NSC mass versus galactic host mass correlation is obtained when GCs have an initial spatial distribution  equal to that of the galaxy, assumed to be in the Dehnen's form. This because the relation shows the same flattening at high galactic masses than observed, while in the case of a different initial density profile for GCs, the $M_{NSC}-M_g$ relation keeps raising at high masses (Sect. \ref{tid});
\item finally, our results suggest that in galaxies with masses above few times $10^{11}$M$_\odot$,
hosting central black holes more massive than $10^8$ M$_\odot$, tidal processes dominate over dynamical friction, leading to NSCs too ``small'' to emerge from the galactic background and be detected. This agrees with both ancient, general, results by 
\cite{Dolc93}, \cite{Tesseri97} and \cite{Tesseri99} and the more specific recent results by \cite{Ant13}.
\end{enumerate}
The overall conclusion is that the migratory-merger model for the formation of dense stellar agglomerates in galactic centers seems to be valid for a large range of types and masses of galaxies giving scaling relations in good agreement with observations and providing a possible explanation for the lack of
NSCs in bright galaxies. An important topic which remains to be investigated thoroughly is what fraction of young to old stars actually reside in NSC, thing which can constitute an important test for the NSC formation models.

\section*{Acknowledgments}

We thank the anonymous referee, whose comments and suggestions allowed us to improve  the paper.

\footnotesize{
\bibliographystyle{mn2e}
\bibliography{bblgrphy}
}

\section*{Appendix}
Considering as density law the $\gamma$ profile in Equation \ref{Mnsct} leads to:

\begin{equation}
N(r)=N\left(\frac{r}{r+R_{\rm g}}\right)^{3-\gamma},
\end{equation}

and the NSC mass is thus given by:
\begin{equation}
M_{\rm NSC}(t)=\Gamma \int_{M_{\rm l}}^{M_{\rm u}}M^{1-s}\left(\frac{CM^{0.38}}{CM^{0.38}+R_{\rm g}}\right)^{3-\gamma}\mathrm{d}M,
\end{equation}

being $C=C(M_{\rm g},R_{\rm g},\gamma,e;t)$ as in Equation \ref{MNSC}. 

The explicit expression in this case is given by:

\begin{equation}
M_{\rm NSC}=
\frac{\mathcal{F}\left[-M_{\rm u}^k\tensor[_2]{F}{_1}\left(a,b;c;z(M_{\rm u})\right)
+M_{\rm l}^k\tensor[_2]{F}{_1}\left(a,b;c;z(M_{\rm l})\right)\right]
}{R_{\rm g}^{3-\gamma}(0.67(\gamma-3)+1.76(s-2))},
\label{sol}
\end{equation}

with $\mathcal{F}=1.76\Gamma_0 C^{3-\gamma}/\rho_0$, $k=0.38(3-\gamma)+2-s$ and $\tensor[_2]{F}{_1}(a,b;c;z)$ the Gauss' Hypergeometric Function, defined as:
\begin{equation}
\tensor[_2]{F}{_1}(a,b;c;z)=\frac{\Gamma(c)}{\Gamma(b)\Gamma(b-c)}\int_0^1 \frac{t^{b-1}(1-t)^{c-b-1}}{(1-tz)^a}dt,
\end{equation}

where $\Gamma(x)$ is the classic Euler's Gamma function \citep{AbSt}

and the arguments in Equation \ref{sol} are:
\begin{eqnarray} \nonumber
a&=&3-\gamma, \\ \nonumber
b&=&-\frac{0.67(\gamma-3)+1.76(s-2)}{0.67}, \\ \nonumber
c&=&-\frac{0.67(\gamma-4)+1.76(s-2)}{0.67}, \\ \nonumber
z&=&-\frac{CM^{0.38}}{R_{\rm g}} .\\ \nonumber 
\end{eqnarray}

\end{document}